\documentclass[graybox, nosecnum]{svmult}

\usepackage{mathptmx}       
\usepackage{helvet}         
\usepackage{courier}        
\usepackage{type1cm}        
\usepackage{amsmath}
\usepackage{makeidx}         
\usepackage{graphicx}        
\usepackage{multicol}        
\usepackage[bottom]{footmisc}
\usepackage{hyperref}        
\usepackage{soul}            
\usepackage{bbold}

\makeindex             

\begin{document}
\title*{Quantum microscopic dynamical approaches}
\author{C\'edric Simenel \thanks{corresponding author}}
\institute{C\'edric Simenel \at Department of Fundamental and Theoretical Physics, Research School of Physics, The Australian National University, Canberra ACT  2601, Australia, \email{cedric.simenel@anu.edu.au}}
%
%
\maketitle
\abstract{Nuclear physics is ideal to test and develop techniques to describe the microscopic dynamics of  quantum many-body systems. At low energy, nuclear dynamics is described with non-relativistic approaches based on the mean-field approximation and its extensions. 
Variational principles based on the stationarity of the action are introduced to build theoretical models with different levels of approximation. In particular, the time-dependent Hartree-Fock (TDHF) equation for mean-field dynamics and its linear approximation, also known as the Random Phase Approximation (RPA), are derived. Predictions of vibrational spectra at the RPA level are presented as an application.
The inclusion of beyond TDHF correlations and fluctuations are then discussed. In particular, pairing correlations are treated at the BCS and Bogoliubov levels.  The Balian-V\'en\'eroni variational principle is finally introduced. In addition to provide some insight into mean-field limitations, 
it offers a possibility to incorporate quantum fluctuations of one-body observables with the time-dependent RPA formalism. }

\section{\textit{Introduction}}

The goal of this chapter is to introduce tools to describe the dynamics of nuclear systems out of equilibrium through their explicit time-evolution. 
This  allows  to investigate nuclear responses to external excitations leading to collective motion such as vibration, rotation and fission. 
The same tools are also used to study heavy-ion collisions such as fusion and (multi)nucleon transfer reactions. 

Our focus in on the tools and concepts rather than applications. 
The ambition, here, is not to provide a review with an exhaustive list of references. 
The methods discussed in this Chapter, as well as their applications to nuclear dynamics have been recently reviewed in (Simenel 2012; Lacroix and Ayik 2014; Nakatsukasa et al 2016, Simenel and Umar 2018; Sekizawa 2019; Stevenson and Barton 2019), which the reader is invited to consult for further details. 
See also the  Sky3D solver (Maruhn et al 2014; Schuetrumpf et al 2018).

Our framework is limited to low-energy nuclear dynamics, with typically few MeV per nucleon. 
In this case, the internal structure of nucleons in terms of quark and gluon degrees of freedom can be neglected. 
The nucleons are approximated as point-like spin $1/2$ fermions without internal excitations. 
They also carry an isospin $1/2$ to distinguish between protons and neutrons. 
Another consequence of the limitation to low-energy dynamics is that relativistic effects can be neglected in a first approximation.

The nuclei are then treated as non-relativistic quantum many-body systems, with up to $\sim500$ nucleons in the case of actinide collisions (the heaviest nuclear systems that can be studied on Earth). 
Ideally, one has to solve the time-dependent Schr\"odinger equation 
\begin{equation}
i\hbar\frac{d}{dt}|\Psi(t)\rangle=\left[\sum_{i=1}^A\frac{\hat{p}(i)^2}{2m}+\sum_{i>j=1}^{A}\hat{v}(i,j)\right]\,|\Psi(t)\rangle,
\label{eq:Schrodinger}
\end{equation}
where $|\Psi(t)\rangle$ is the time-dependent many-body quantum state and $\hat{v}(1,2)$ the interaction between 2 nucleons. 
(For simplicity, 3-body interactions and higher are omitted.)

There are two main problems here. 
The first one is that $\hat{v}(1,2)$ is not well known. 
Unfortunately, this interaction is different to the one between two isolated nucleons (which can be measured through scattering experiments), and connecting the interaction in vacuum with the interaction in medium is not straightforward. 
This is in part due to the fact that the nucleons are 
 not point like objects and could be polarised by surrounding nucleons, thus modifying the interaction in a non-straightforward way. 
Another difficulty is that relativity leaves significant traces in the nuclear interaction such as spin-orbit interaction terms. 

The second problem is that it is usually impossible to solve Eq.~(\ref{eq:Schrodinger}) exactly. 
As a result, approximations to the quantum many-body problem are needed.
A compromise is often needed between the complexity and level of precision of the mechanism one wants to describe in one hand, and, in the other hand, the computational capabilities that are available. 
This is the problem that is addressed here, leaving the very interesting and highly challenging issue with the in-medium interaction to others. 
It is  thus assumed that  $\hat{v}(1,2)$ is ``known'', and the focus of this chapter on building efficient approximations to the time-dependent quantum many-body problem to describe aspects of low-energy nuclear dynamics. 

As a reminder (as well as an introduction to the notation), this chapter starts with a description of many-body states.
The question ``Why can't we solve the time-dependent Schr\"odinger equation exactly for many-body systems?'' is then answered.
Variational principles are introduced as a tool to optimise many-body dynamics.
The time-dependent Hartree-Fock (TDHF) theory and its linearisation leading to the Random Phase (RPA) approximation are described in the following two sections. 
The inclusion of pairing correlations are then done within the time-dependent Hartree-Fock-Bogoliubov (TDHFB) and BCS approaches.
The Balian-V\'en\'eroni variational principle and its application to one-body fluctuations through the time-dependent RPA are discussed in the last section.

\section{Many-body states}
\label{sec:many-body}

\subsection{One-particle states}

The quantum state of a single-particle is noted $|\varphi\rangle$. It belongs to the Hilbert space of one particle, noted $\mathcal{H}_1$.
In the framework of second quantisation, such a state is written $|\varphi\rangle=\hat{a}^\dagger_\varphi|-\rangle$, where $|-\rangle$ is the state of the vacuum and $\hat{a}^\dagger_\varphi$ creates a particle in the state $|\varphi\rangle$.

\subsection{Two distinguishable particles}

When two particles are distinguishable, such as a proton and an electron in the Hydrogen atom, or a proton and a neutron in a deuteron, the state of each particle can be labelled, e.g., with a number. 
Two cases are considered: one where the particles are independent and one where they are correlated. 

\subsubsection{Independent particles}

The state of two independent particles can be written as $|\Phi\rangle=|1:\varphi_1,2:\varphi_2\rangle$. 
In this case, the state of one particle does not depend on the state of the other particle. 
The notations $|1:\varphi_1,2:\varphi_2\rangle =|\varphi_1\rangle\otimes|\varphi_2\rangle =|\varphi_1\rangle|\varphi_2\rangle$ are used equivalently. 

\subsubsection{Correlated particles}

The state of two correlated particles is written as a sum of  independent particle states: 
$|\Psi\rangle=\sum_\alpha C_\alpha|1:\varphi_\alpha,2:\varphi'_{\alpha}\rangle$.
The correlation reads as follows:
\begin{itemize}
\item If particle 1 is in $\varphi_a$, then particle 2 is in $\varphi_a'$
\item If particle 1 is in $\varphi_b$, then particle 2 is in $\varphi_b'$
\item $\cdots$
\end{itemize}
The coefficients $C_\alpha$ are the amplitude of probability for each configuration. 

For example, in the case of a deuteron, if a proton is found in $\mathbf{r}$, the neutron is  at a nearby position and not elsewhere.  
In the case of a prolately deformed nucleus, if one nucleon is found at one tip of the nucleus, then another one has to be present at the other tip.

\subsection{Two indistinguishable fermions}

Now consider the case of two identical fermions which are indistinguishable. 
In this case, the states $|1:\varphi_1,2:\varphi_2\rangle$ and $|1:\varphi_2,2:\varphi_1\rangle$ are equivalent and must both appear with the same probability in the two-body state. 
For two independent fermions, 
$$|\Phi\rangle=|\varphi_1\varphi_2\rangle=(|1:\varphi_1,2:\varphi_2\rangle-|1:\varphi_2,2:\varphi_1\rangle)/\sqrt{2}.$$
The minus sign accounts for the fact that the state must be antisymmetric for fermions (there would be a plus sign for bosons). 
The antisymmetry is obvious as $|\varphi_1\varphi_2\rangle=-|\varphi_2\varphi_1\rangle$. 
In particular, $|\varphi\varphi\rangle=0$ which is a manifestation of the Pauli principle. 

The state $|\Phi\rangle$ can be written in the form of a Slater determinant
$$|\Phi\rangle=\frac{1}{\sqrt{2}}\,
\begin{vmatrix}
|1:\varphi_1\rangle & |1:\varphi_2\rangle \\
|2:\varphi_1\rangle & |2:\varphi_2\rangle \\
\end{vmatrix}
=\mathcal{A}|1:\varphi_1,2:\varphi_2\rangle,
$$
where $\mathcal{A}$ is the antisymmetrisation operator. 

In second quantisation, $|\Phi\rangle=\hat{a}^\dagger_2\hat{a}^\dagger_1|-\rangle$, with the antisymmetry ensured by the anticommutation relationship $\{{\hat{a}^\dagger}_1,{\hat{a}^\dagger}_2\}=0$.
The generalisation to correlated two-particle states becomes 
$|\Psi\rangle=\sum_\alpha C_\alpha\hat{a}^\dagger_{\alpha_2}\hat{a}^\dagger_{\alpha_1}|-\rangle$.
Both $|\Phi\rangle$ and $|\Psi\rangle$ belong to the Hilbert space of two identical fermions $\mathcal{H}_2$.

\subsection{Many-fermion states}

The generalisation to the case of $N$ identical fermions is now straightforward.
For independent particles, the many-body state is described by a Slater determinant
$$|\Phi\rangle=\frac{1}{\sqrt{N!}}\,
\begin{vmatrix}
|1:\varphi_1\rangle &\cdots & |1:\varphi_N\rangle \\
\vdots&&\vdots\\
|N:\varphi_1\rangle&\cdots & |N:\varphi_N\rangle \\
\end{vmatrix}
=\mathcal{A}|1:\varphi_1, \cdots,N:\varphi_N\rangle=\hat{a}^\dagger_N\cdots\hat{a}^\dagger_1|-\rangle.
$$

As before, a correlated state is written as a sum of independent particle states $|\Psi\rangle=\sum_\alpha C_\alpha|\Phi_\alpha\rangle$.
As a result, a complete set of orthonormal Slater determinants constitute a possible basis of the Hilbert space of $N$ particles $\mathcal{H}_N$.

\section{Why can't  the many-body time-dependent Schr\"odinger equation be solved exactly?}
\label{sec:schrodinger}

Consider a Slater (uncorrelated many-body state) $|\Psi(t_0)\rangle=|\Phi_0\rangle$ at initial time  $t_0$.
The goal is to evaluate the state at a later time $t_1$.
This can be done using the evolution operator 
$$|\Psi(t_1)\rangle=e^{-i\hat{H}(t_1-t_0)/\hbar}|\Phi_0\rangle,$$
with the Hamiltonian
$$\hat{H}=\sum_{i=1}^A\frac{\hat{p}(i)^2}{2m}+\frac{1}{2}\sum_{i,j=1}^A \hat{v}(i,j).$$
The first term of the Hamiltonian (the kinetic energy) is a one-body operator, while the second term (the interaction) is a two-body operator.
 (The $\frac{1}{2}$ factor is to avoid double counting interactions between pairs of particles.)
 
 If $\hat{v}=0$ (no interaction), then the particles remain independent and evolve according to their own kinetic energy operator $\frac{\hat{p}^2}{2m}$. 
In this case, the state remains a single Slater determinant and the problem becomes trivial. Of course, the difficulty comes from the interaction when it is non-zero.

The exponential of an operator is usually expressed as a Taylor expansion. 
As the latter is an infinite sum,  it needs to be truncated.
For a time evolution operator, this is done  with  a small time increment $\Delta t$.
The evolution over $\Delta t$ is then repeated many times to reach the desired time $t_1$.
$$|\Psi(t_1)\rangle=\cdots e^{-i\hat{H}\Delta t/\hbar} \cdots|\Phi_0\rangle.$$
Choosing $\Delta t$ small enough, the exponential can be approximated in the first order, $e^{-i\hat{H}\Delta t/\hbar}\simeq1-i\hat{H}\Delta t/\hbar$. 

The first iteration reads $|\Psi(t_0+\Delta t)\rangle\simeq|\Phi_0\rangle-i\hat{H}\Delta t |\Phi_0\rangle/\hbar$. 
One then has  to compute the action of the interaction on the initial Slater, $\sum_{i,j=1}^A \hat{v}(i,j)|\Phi_0\rangle$. 

Let us use some simple (and maybe not so rigorous) arguments to evaluate the complexity of the problem.
If there was only one sum $\sum_i$ instead of two ($\sum_{i,j}$), then the problem would be mathematically similar to the $\hat{v}=0$ case, i.e., the Slater would remain a Slater at all times.
However, there is an additional sum $\sum_{j=1}^A$ which means that the state at $t_0+\Delta t$ is a sum of $A$ (the number of particles) Slaters.
One then expects $A^2$ Slaters at $t_0+2\Delta t$ and, more generally, $A^n$ Slaters at $t_0+n\Delta t$.
This exponential increase in not tractable. As a result, the many-body Schr\"odinger equation can usually not be solved brut force, and  an approach which allows  to build various levels of approximation is required.

\section{Variational principles}
\label{sec:variational}

Neglecting relativistic effects, the exact evolution is supposed to be given by the time-dependent Schr\"odinger equation. 
Thus, any theory that is equivalent to Schr\"odinger when no approximations are made is as ``good'' as Schr\"odinger. 

\subsection{Variational principle with the Dirac action}

There are several variational principles that could be used. For the moment,  consider a variational principle based on Dirac's action 
$$S=\int_0^T dt\,\langle\Psi|\left(i\hbar\frac{d}{dt}-\hat{H}\right)|\Psi\rangle.$$
Requiring the stationarity of the action, $\delta S=0$, leads to the Schr\"odinger equation. 

Remember that, for a complex variable $z={\mathrm{Re}}(z)+i\,{\mathrm{Im}}(z)$, there are two independent variables, ${\mathrm{Re}}(z)$ and ${\mathrm{Im}}(z)$, or, equivalently, $z={\mathrm{Re}}(z)+i\,{\mathrm{Im}}(z)$ and $z^*={\mathrm{Re}}(z)-i\,{\mathrm{Im}}(z)$.
Similarly, $|\Psi\rangle$ and $\langle\Psi|$ can be treated as independent variational quantities and  the property $\langle\Psi|=(|\Psi\rangle)^\dagger$ can be restored at the end.
The stationarity condition is then expressed as 
$$\,\,\,\frac{\delta S}{\delta \langle\Psi|}=0\,\,\,\mbox{ and }\frac{\delta S}{\delta |\Psi\rangle}=0  \,\,\,\mbox{ for }\,\,\, 0\le t\le T.$$

The first condition gives
\begin{eqnarray}
\frac{\delta }{\delta \langle\Psi(t)|}\int_0^T dt'\,\langle\Psi(t')|\left(i\hbar\frac{d}{dt}-\hat{H}\right)|\Psi(t')\rangle&=&0\nonumber\\
\int_0^T dt'\,\frac{\delta \langle\Psi(t')| }{\delta \langle\Psi(t)|}\left(i\hbar\frac{d}{dt}-\hat{H}\right)|\Psi(t')\rangle&=&0\nonumber\\
\left(i\hbar\frac{d}{dt}-\hat{H}\right)|\Psi\rangle&=&0\nonumber
\end{eqnarray}
where  $\frac{\delta \langle\Psi(t')| }{\delta \langle\Psi(t)|}=\delta(t-t')$ was used. The last line is the Schr\"odinger equation. 

The second condition can be expressed as 
$$\frac{\delta }{\delta|\Psi(t)\rangle}\int_0^T dt'\,\langle\Psi(t')|\left(i\hbar\frac{d}{dt}-\hat{H}\right)|\Psi(t')\rangle=0.$$
One has to be careful with the $\frac{d}{dt}$ term. 
Using integration by part, 
\begin{eqnarray}
\frac{\delta }{\delta|\Psi(t)\rangle}\int_0^T dt'\,\langle\Psi(t')|\left(\frac{d}{dt}|\Psi(t')\rangle\right)&=&\nonumber\\
\frac{\delta }{\delta|\Psi(t)\rangle}\left[ \left[\langle\Psi(t')|\Psi(t')\rangle\right]_0^T-\int_0^T dt'\,\left(\frac{d}{dt}\langle\Psi(t')|\right)|\Psi(t')\rangle\right]&=&\nonumber\\
\left[\langle\Psi(t')|\frac{\delta |\Psi(t')\rangle }{\delta|\Psi(t)\rangle}\right]_0^T-\int_0^T dt'\,\left(\frac{d}{dt}\langle\Psi(t')|\right)\frac{\delta |\Psi(t')\rangle }{\delta|\Psi(t)\rangle} &=&\nonumber\\
\left[\langle\Psi(t')|\frac{\delta |\Psi(t')\rangle }{\delta|\Psi(t)\rangle}\right]_0^T-\frac{d}{dt}\langle\Psi(t)|.&&\label{eq:dtterm}
\end{eqnarray}
(Note that  $\frac{\delta \langle\Psi(t')| }{\delta \langle\Psi(t)|}=\delta(t-t')$ in the $[\cdots]_0^T$ term was not used intentionally.)

The term with $\hat{H}$ is trivial if the latter does not depend on $|\Psi\rangle$:
\begin{equation}
\frac{\delta }{\delta|\Psi(t)\rangle}\int_0^T dt'\,\langle\Psi(t')|\hat{H}|\Psi(t')\rangle=\langle\Psi(t)|\hat{H}.\label{eq:Hterm}
\end{equation}
Combining Eqs.~(\ref{eq:dtterm}) and~(\ref{eq:Hterm}), 
\begin{equation}
i\hbar\left(\langle\Psi(T)|\frac{\delta |\Psi(T)\rangle }{\delta|\Psi(t)\rangle}-
\langle\Psi(0)|\frac{\delta |\Psi(0)\rangle }{\delta|\Psi(t)\rangle}
-\frac{d}{dt}\langle\Psi(t)|\right)-\langle\Psi(t)|\hat{H}=0.
\label{eq:term2}
\end{equation}

Now ``restore'' the property $\langle\Psi|=(|\Psi\rangle)^\dagger$.
As shown earlier, imposing $\frac{\delta S}{\delta \langle\Psi|}=0$ leads to the Schr\"odinger equation. 
Taking its Hermitian conjugate, 
$-i\hbar\frac{d}{dt}\langle\Psi|=\langle\Psi|\hat{H}$. 
Comparing with Eq.~(\ref{eq:term2}), it is seen that, to have $\frac{\delta S}{\delta |\Psi\rangle}=0$, one needs to have
$$\langle\Psi(T)|\frac{\delta |\Psi(T)\rangle }{\delta|\Psi(t)\rangle}-
\langle\Psi(0)|\frac{\delta |\Psi(0)\rangle }{\delta|\Psi(t)\rangle}=0.$$
This is obtained by forbidding variations at the initial and final times, i.e., $\delta|\Psi(0)\rangle=\delta|\Psi(T)\rangle=0$.

To summarise, defining the Dirac action $S=\int_0^T dt\,\langle\Psi|\left(i\hbar\frac{d}{dt}-\hat{H}\right)|\Psi\rangle$ and 
requiring $\frac{\delta S}{\delta \langle\Psi|}=0$ leads to Schr\"odinger equation, while $\frac{\delta S}{\delta |\Psi\rangle}=0$ gives its Hermitian conjugate {\it if} it is  required that $\Psi$ cannot vary at $t=0$ and $t=T$. 

\subsection{Variational space}

To get Schr\"odinger from the variational principle $\delta S=0$, the variational space for $\Psi$ was not restricted, i.e., it spans the entire Hilbert space $\mathcal{H}_N$. 
If  the variational space was restricted to a sub-space $\mathcal{F}_N$ of $\mathcal{H}_N$, one would expect to get a solution to the variational principle $|\Psi(t)\rangle$ that is not solution to the Schr\"odinger equation, but which is an optimised approximation of the exact evolution within $\mathcal{F}_N$. 
The role of a theorist  is then to choose a sub-space $\mathcal{F}_N$ which  {\it (i)} contains enough physics to give a good approximation to the evolution of the system of interest, and {\it(ii)} is simple enough so that the resulting equations of motion can be solved analytically or numerically.

\section{Time-Dependent Hartree-Fock (TDHF) theory}
\label{sec:TDHF}

If the state of the system was
to remain a Slater at all time, then the evolution
would be in principle easier to evaluate.
The TDHF theory, as proposed by Dirac in 1930, is built
upon the approximation that the state of the many-body system remains in the sub-space $\mathcal{F}_N$ of
Slater determinants. To get the TDHF equation,
one then needs to solve the variational principle
$\delta S=0$  within this subspace. 

\subsection{TDHF equation}

The Dirac action is 
$$S_{t_0,t_1}[\Phi]=\int_{t_0}^{t_1} dt\,\langle\Phi(t)|\left(i\hbar\frac{d}{dt}-\hat{H}\right)|\Phi(t)\rangle$$
where $|\Phi\rangle\in\mathcal{F}_N$ is a Slater determinant built from the $A$ occupied single-particle states $|\varphi_i\rangle$, $1\le i\le A$. 

The expectation value of the time derivative is computed first. From 
$$\frac{d}{dt}|\Phi\rangle=\frac{d}{dt}\mathcal{A}|\varphi_A\rangle\cdots|\varphi_1\rangle=
\mathcal{A}\sum_{j=1}^A|\varphi_A\rangle\cdots\left(\frac{d}{dt}|\varphi_j\rangle\right)\cdots|\varphi_1\rangle$$
one gets
\begin{eqnarray}
\langle\Phi|\frac{d}{dt}|\Phi\rangle &=& 
\left[\langle\varphi_1|\cdots\langle\varphi_A| \mathcal{A} \right]
\mathcal{A}\sum_{j=1}^A|\varphi_A\rangle\cdots\left(\frac{d}{dt}|\varphi_j\rangle\right)\cdots|\varphi_1\rangle\nonumber\\
&=& \sum_{j=1}^A\langle\varphi_1|\varphi_1\rangle\cdots\left(\langle\varphi_j|\frac{d}{dt}|\varphi_j\rangle\right)\cdots\langle\varphi_A|\varphi_A\rangle\nonumber\\
&=& \sum_{j=1}^A \langle\varphi_j|\frac{d}{dt}|\varphi_j\rangle.\nonumber
\end{eqnarray}

Now compute the expectation value of the Hamiltonian and write it as an energy density functional (EDF) $E[\rho]$:
$$\langle\Phi|\hat{H}|\Phi\rangle=\langle\varphi_1\cdots\varphi_A|\hat{H}|\varphi_A\cdots\varphi_1\rangle=E[\varphi_1\cdots\varphi_A]\equiv E[\rho],$$
where $\rho$ is the one-body density matrix. 
Its elements are defined as 
\begin{equation}
\rho_{\alpha\beta}=\langle\hat{a}_\beta^\dagger\hat{a}_\alpha\rangle.\label{eq:defrho}
\end{equation}
For a Slater, 
\begin{equation}
\rho_{\alpha\beta}=\langle\varphi_1\cdots\varphi_A|\hat{a}^\dagger_\beta\hat{a}_\alpha|\varphi_A\cdots\varphi_1\rangle
=\sum_{j=1}^A\langle\varphi_j|\beta\rangle\langle\alpha|\varphi_j\rangle.\label{eq:rhoSlater}
\end{equation}
For instance, in the position-spin-isospin basis of $\mathcal{H}_1$, $\{|\mathbf{r},s,q\rangle\}$, one has
$$\rho(\mathbf{r}sq,\mathbf{r}'s'q')=\sum_{j=1}^A\varphi_j^*(\mathbf{r}'s'q')\varphi_j(\mathbf{r}sq).$$
The one-body density matrix  is expressed as a function of all the occupied single-particle states.
It contains the same information on the system as the Slater. 
This is why  the functional $E[\varphi_1\cdots\varphi_A]$ can be replaced by the energy density functional $E[\rho]$.
The EDF accounts for the kinetic energy and the interaction between the particles.
The most popular EDF are the  Skyrme and Gogny functionals [see (Bender et al 2003) for a review].

The action is then written as 
\begin{eqnarray}
S &=& \int_{t_0}^{t_1}  d t \, \left( i\hbar\sum_{i=1}^N \langle\varphi_i| \frac{ d}{ d t} |{\varphi}_i\rangle
-E[\rho(t)]\right) \nonumber \\
&=& \int_{t_0}^{t_1}   d t \, \left(i\hbar \sum_{i=1}^N  \, \int   d x \,  \varphi_i^*(x,t) \, \frac{ d}{ d t} {\varphi}_i(x,t)
-E[\rho(t)]\right) \nonumber \\
\end{eqnarray}
where   the notations $x\equiv (\mathbf{r} s q)$ and $\int  d x = \sum_{q s} \int   d \mathbf{r}$ are introduced.

The variational principle  $\delta S = 0$ is solved by considering the variation of all the independent variables 
$\varphi_i$ and $\varphi^*_j$, $1\le j\le A$:
\begin{eqnarray}
\frac{\delta S}{\delta  \varphi_j(x,t)} = 0 \mbox{\hspace{1cm} and \hspace{1cm}}
\frac{\delta S}{\delta  \varphi^*_j(x,t)} = 0 .
\label{eq:delphiphistar}
\end{eqnarray}

The variation over $\varphi^*$  gives
\begin{equation}
\frac{\delta S}{\delta  \varphi^*_j(x,t)} = i\hbar \, \frac{ d}{ d t} \varphi_j(x,t) - 
 \int_{t_0}^{t_1}   d t' \, \frac{\delta E[\rho(t')]}{\delta  \varphi^*_j(x,t)}.
\end{equation}
The functional derivative of $E$ can be re-written thanks to a change of variable
\begin{equation} 
 \frac{\delta E[\rho(t')]}{\delta  \varphi^*_j(x,t)}
 =  \int  d y \,  d y' \, \frac{\delta E[\rho(t')]}{\delta  \rho(y,y';t')}
\, \frac{\delta  \rho(y,y';t')}{\delta  \varphi^*_j(x,t)}.
\end{equation}
Using
\begin{equation}
  \frac{\delta  \rho(y,y';t')}{\delta  \varphi^*_j(x,t)} =  \varphi_j(y,t') \, \delta(y'-x) \, \delta(t-t')
\end{equation}
and noting the single-particle Hartree-Fock Hamiltonian $h$ with matrix elements
\begin{equation}
h(x,y;t) = \frac{\delta E[\rho(t)]}{\delta \rho(y,x;t)},
\label{eq:hEHF}
\end{equation}
one gets the TDHF equation for the set of occupied states
\begin{equation}
\fbox{$ \displaystyle i\hbar \, \frac{ d }{ d t} \varphi_j(x,t) = \int   d y \, h(x,y;t) \, \varphi_j(y,t)$}. \label{eq:TDHF_az}
\end{equation}
As in the Schr\"odinger case, the variation over $\varphi$ leads to the complex conjugate of the TDHF equation. 

The TDHF equation can also be written in a matrix form
$$i\hbar\frac{d}{dt}\varphi_j(t)=h(t)\varphi_j(t) \,\,, \,\,\,\,\,1\le j\le A,$$
or, using Dirac notation,
\begin{equation}
i\hbar\frac{d}{dt}|\varphi_j(t)\rangle=\hat{h}(t)|\varphi_j(t)\rangle.\label{eq:TDHFDirac}
\end{equation}
Note that the single-particle Hamiltonian $h$ is self-consistent, i.e., it depends on the density matrix $\rho$.
It is then also a function of time.  
As a result, the TDHF equation is equivalent to a set of non-linear single-particle Schr\"odinger equations. 
The non-linearity comes from the self-consistency of $h[\rho]$: 
The Hamiltonian is a function of the states on which it acts. 

\subsection{Liouville form of the TDHF equation}

The one-body density matrix can be written as an operator of $\mathcal{H}_1$: 
$$\hat{\rho}=\sum_{j=1}^A|\varphi_j\rangle\langle\varphi_j|.$$
Indeed, its matrix elements 
$$\rho_{\alpha\beta}=\langle\alpha|\hat{\rho}|\beta\rangle=\sum_{j=1}^A\langle\varphi_j|\beta\rangle\langle\alpha|\varphi_j\rangle$$
are the same as for a Slater determinant in Eq.~(\ref{eq:rhoSlater}).

Multiplying Eq.~(\ref{eq:TDHFDirac}) by $\langle\varphi_j|$ on the right and summing over $j$, 
\begin{equation}
\sum_{j=1}^Ai\hbar\left(\frac{d}{dt}|\varphi_j\rangle\right)\langle\varphi_j|=\sum_{j=1}^A\hat{h}|\varphi_j\rangle\langle\varphi_j|.\label{eq:TDHFket}
\end{equation}
Similarly,  multiplying the Hermitian conjugate of Eq.~(\ref{eq:TDHFDirac}) by $|\varphi_j\rangle$ on the left and summing over $j$ leads to
\begin{equation}
\sum_{j=1}^A-i\hbar|\varphi_j\rangle\left(\frac{d}{dt}\langle\varphi_j|\right)=\sum_{j=1}^A|\varphi_j\rangle\langle\varphi_j|\hat{h}.\label{eq:TDHFbra}
\end{equation}
Taking the difference between Eqs.~(\ref{eq:TDHFket}) and~(\ref{eq:TDHFbra}) gives the Liouville form of the TDHF equation
\begin{eqnarray}
\sum_{j=1}^Ai\hbar\frac{d}{dt}\left(|\varphi_j\rangle\langle\varphi_j|\right)&=&\sum_{j=1}^A\left(\hat{h}|\varphi_j\rangle\langle\varphi_j|-|\varphi_j\rangle\langle\varphi_j|\hat{h}\right)\nonumber\\
i\hbar\frac{d}{dt}\hat{\rho}&=& \left[\hat{h},\hat{\rho}\right].\label{eq:TDHFLiouvillehat}
\end{eqnarray}
This can be written also in the matrix form
\begin{equation}
i\hbar\frac{d}{dt}{\rho}= \left[{h},{\rho}\right].
\end{equation}

\subsection{Solving TDHF}

The easiest way to solve the TDHF equation numerically is
to start from the non-linear Schr\"odinger equation for occupied
single-particles wave-functions
$$i\hbar\frac{d}{dt}\varphi_j(\mathbf{r}sq,t)=\sum_{s'q'}\int d^3r' \, h(\mathbf{r}sq,\mathbf{r}'s'q';t)\,\varphi_j(\mathbf{r}'s'q',t)\,,\,\,\,\,\,1\le j\le A.$$
Because $h\equiv h[\rho(t)]$ is time-dependent, one cannot use $e^{-ih(t_1-t_0)/\hbar}$ as an evolution operator,  unless one considers an evolution over a time interval that is short enough so that the  variation of  $h(t)$  during this  time interval can be neglected, i.e., 
$\varphi_j(t+\Delta t)\simeq e^{-ih\Delta t/\hbar}\varphi_j(t)$. 

To ensure that the evolution is unitary (required for energy
and particle number conservations), one needs the same operator
(up to Hermitian conjugation) to go from $t\rightarrow t+\Delta t$
as the one to
do the time-reversed operation $t +\Delta t \rightarrow t$. 
Therefore, $h$ has to be estimated at $t+\frac{\Delta t}{2}$. 

A possible algorithm is 
\begin{equation}
\begin{array}{ccccc}
\{ \varphi_j(t) \} &\!\!\!\!\! \!\!\!\!\!\!\!\!\!\!\Rightarrow &\rho(t) &\Rightarrow & h(t) \\
\Uparrow &&&& \Downarrow\\
{\varphi}_i(t+\Delta t)  =  e^{-i\frac{\Delta t}{\hbar}  h(t+\frac{\Delta t}{2}) }{\varphi}_i(t)
&&&&\!\!\!\!\!\!\!\!\!\!\tilde{\varphi}_i(t+\Delta t)  =  e^{-i\frac{\Delta t}{\hbar}  h(t)} {\varphi}_i(t)  \\
\Uparrow &&&& \Downarrow\\
h(t+\frac{\Delta t}{2})&\!\!\!\!\!\!\!\!\!\!\!\!\!\!\Leftarrow&{\rho}(t+\frac{\Delta t}{2})=\frac{\tilde{\rho}(t+\Delta t)+{\rho}(t)}{2} &\Leftarrow&\tilde{\rho}(t+\Delta t)\end{array} 
\label{eq:algo}
\end{equation}
The initial state $\{ \varphi_j(t_0) \}$ could be a Hartree-Fock (HF) ground-stale
which is put into motion thanks to an external one-body time
dependent potential added to the mean-field. 
For collisions, one usually starts with two non-overlapping HF ground-states at some
finite distance in a large box and a Galilean
boost $e^{i\mathbf{k}_n\cdot\mathbf{r}}$ ($n=1,2$ denotes the two nuclei) is applied to
induce a momentum $\hbar \mathbf{k}_n$ to the nucleons at the initial
time $t_0$.

Typical TDHF solvers use cartesian grids with mesh spacing $\Delta x=0.6-1.0$~fm.
Several types of boundary conditions can be used: 
\begin{itemize}
\item {\it Hard} (particles are reflected by an infinite potential on the edges of the box)
\item {\it Periodic} (particles reaching one edge reappear on the other side) 
\item {\it Absorbing} (particles are absorbed on the edge)
\end{itemize} 
Absorbing boundary conditions can be obtained with an additional layer of imaginary potential outside the box, or by ``twisting'' the phase on the boundaries (Schuetrumpf et al 2016).

\subsection{Static HF}

Static HF ground-states are required  to build
the TDHF initial condition. The static HF equation
reads 
$$\left[h[\rho],\rho\right]=0.$$
As they commute, one can find a
single-particle states basis that diagonalises both $\rho$ and $h$ simultaneously.
(Note that to solve TDHF, 
a basis needs to be chosen. TDHF equations are usually solved in the ``canonical'' basis in which $\rho$
is diagonal. In this basis, $h$ is not diagonal 
if $h$ and $\rho$ do not commute, i.e., when the state evolves in time: $[h,\rho]=i\hbar \dot\rho$.)

In principle the HF ground-states could be determined from any HF code as long as the same energy density functional is used as in the TDHF evolution. However, to minimise numerical errors, it is best to use a HF code that uses the same numerical approximations as in TDHF, in particular in term of mesh grid and spatial derivatives. 

One way to do this is by solving the HF equation with the imaginary time method, i.e.,
replacing $t\rightarrow- i\tau$ in the TDHF equation. This makes
an initial state (usually built from Nilsson or harmonic
oscillator wave-functions) converge towards the ground-state.

Of course, in the Hartree-Fock theory, the Hamiltonian is self-consistent
so an iterative process with imaginary time step $\Delta \tau$ is needed.
In addition, $e^{-h\tau/\hbar}$ is not unitary so an 
orthonormalisation of the occupied single-particle wave-functions is required, e.g., with a Gram-Schmidt procedure.

\subsection{Examples of TDHF applications}

Here, some applications are only briefly mentioned. See Reviews 
(Simenel 2012; Lacroix and Ayik 2014; Nakatsukasa et al 2016, Simenel and Umar 2018; Sekizawa 2019; Stevenson and Barton 2019)
for more detailed examples. 

TDHF calculations are now standard to investigate various aspects of nuclear dynamics. 
Collective vibrations can be studied by computing the time evolution of multipole moments of a nucleus following a collective boost excitation (see next section on RPA). 
Although most applications have been dedicated to giant resonances, low-lying collective vibrations have been also studied. 

Fusion barriers can be obtained in a straightforward manner by searching for fusion thresholds in head-on collisions. 
Below this threshold, the two fragments re-separate after contact. 
Once this threshold is overcome, the fragments merge in a single one. 
Compared to barrier heights of the ``bare'' nucleus-nucleus potential, this fusion threshold is often found at a lower energy due to dynamical effects such as vibration and transfer in the entrance channel. 
Applying the same method for searching for fusion thresholds at finite impact parameters allow the computation of above barrier fusion cross-sections. 

Even if the nuclei do not fuse after contact, part of the wave-function can be transferred from one fragment to the other in quasi-elastic collisions. 
The final fragments being a coherent superposition of eigenstates of the particle number operator, one obtains a distribution of probability to find a given number of nucleons in the final fragments. 
Naturally, both fragments are entangled so that ``measuring'' the number of protons or neutrons of one fragment projects the other fragment into a state with the corresponding number of nucleons as the total number of particles is conserved in TDHF. 

Significant efforts have been dedicated in the past decade to study heavy-ion collision at energies well above the barrier, leading to deep inelastic collisions (DIC), as well as in heavy systems leading to quasi-fission where fission-like fragments are produced without the intermediate formation of an equilibrated compound nucleus. 
These reactions offer a wide variety of observables to compare with experiment, such as fragment mass and charge distributions, scattering angle and final kinetic energy. Comparison with theory offers valuable information on expected contact times between the fragments, as well as equilibration and dissipation mechanisms. 

\subsection{Limitations of TDHF}

The TDHF theory is based on the mean-field approximation. 
It describes the evolution of  independent particles in a mean-field potential produced by the ensemble of particles. 
As a result, some correlations are neglected.
This is the case of pairing correlations that lead to the formation of Cooper pairs and induce a superfluid behaviour to some nucleons in the nucleus. 
These correlations are generated by the pairing residual interaction that is neglected in the Hartree-Fock approximation.   

Another limitation is the lack of quantum tunnelling of the many-body wave-function. 
This is due to the restriction of the variational space to a single Slater determinant. 
In reality, colliding nuclei have a non-zero probability to fuse even at sub-barrier energies thanks to tunnelling. 
However, to describe a coherent superposition of a fused system and a system of two outgoing fragments, one would need at least two Slaters, one for the compound nucleus and for the outgoing fragments.
Standard applications of TDHF are then unable to describe sub-barrier fusion or spontaneous fission. 
There are however, techniques such as Density-Constrained TDHF that allow to extract a nucleus-nucleus potential from TDHF simulations of  two colliding nuclei
(Umar and Oberacker 2016). 
Such potential can then be used to evaluate tunnelling probabilities with standard techniques, and then predict fusion cross-sections even below the barrier. 
Alternatively, tunnelling can be studied at the mean-field level through a Wick rotation changing real time into imaginary time (Levit 1980). 
However, this approach has not reached the level of realistic applications yet. 

It is also well known that TDHF lacks of quantum fluctuations. 
This is evident from comparisons between theoretical predictions and experimental measurements of fragment particle number distributions in deep-inelastic collisions in which TDHF underpredicts widths of these distributions. 
This is again a limitation due to the restriction to a single Slater determinant. 
Indeed, in TDHF, one Slater is used to describe all exit channels. 
While the mean-field dynamics of this Slater is expected to give a reasonable description of the most likely outcome, it is unlikely to be realistic for exit channels that deviate significantly from the average one.

\subsection{Justification of the mean-field approximation}

It may seem strange that an independent particle
approximation works for the nucleus which is a very dense
object of nucleons very close to each other and interacting
via the strong and Coulomb interactions.
What makes it work is the Pauli exclusion principle.
The latter prevents the collision of two nucleons to happen
inside the nucleus if the final state of the collision is already occupied. 

Consider two nucleons with an energy $E_1$ and $E_2$ before they interact and an energy $E_1'$ and $E_2'$ after the collision.
Because of energy conservation, and assuming the state of the other nucleons is not changed, one has $E_1+E_2=E_1'+E_2'$.
In the ground state, or a low excitation energy state,
the nucleons have initially an energy lower than the Fermi
energy $E_F$ below which single-particle states are essentially
occupied and above which they are mostly empty. 
This means that $E_1<E_F$ and $E_2<E_F$. 
Thus, even if one final state has $E_i'> E_F$, the other must have
an energy lower than $E_F$ and then one nucleon ends up in a state which
is likely to be already occupied, which is forbidden by Pauli. 
This ``Pauli blocking'' increases the mean-free path of a nucleon to
about the size of the nucleus, reducing the effect of collisions
and thus allowing the mean-field approximation to work.

\section{Random Phase Approximation (RPA)}
\label{sec:RPA}

(The term ``random phase'' refers to another way of deriving the
RPA equations.)

\subsection{Harmonic approximation}

The RPA is used to study small amplitude oscillations of
many-body systems. It is a ``harmonic'' approximation
as in the small amplitude limit (i.e. the system is only
allowed small deformations around the ground-state) the
potential energy varies quadratically, $V\propto(Q-Q_{g.s.})^2$, with
respect to the deformation $Q$ (e.g., multipole moment).
In the harmonic picture, the
relationship between the frequency $\frac{\omega}{2\pi}$ of the
oscillation and the energy spectrum of the harmonic
oscillator $E_n = (n+\frac{1}{2}) \hbar\omega$ can then be used.

\subsection{Transition amplitude}

In addition to the energy $\hbar\omega$ of the
vibration, one is also interested in its collectivity quantified by the
transition amplitude $q_\nu=  \langle\nu| \hat{Q} |0\rangle$
between the ground-state $|0\rangle$ and the first phonon $|\nu\rangle$ of the vibrational spectrum. 
$\hat{Q}=\sum_{i=1}^A \hat{q}(i)$ is a one-body operator, often chosen as a multipole moment
$$\hat{Q}_{LM} =\sum_{i=1}^A \hat{r}^{L+2\delta_{L,0}}Y_{LM}(\hat{\theta},\hat{\phi}).$$

In the expression for the transition amplitude $q_\nu=  \langle\nu| \sum_{i=1}^A \hat{q}(i) |0\rangle$, 
the contributions of each nucleon to the many-body state $|\nu\rangle$ are summed over. Then, the
larger the magnitude of $q_\nu$, the larger the collectivity.
Both $\hbar\omega_\nu$ and $q_\nu$ can be extracted from TDHF using the linear response theory.

\subsection{Linear response theory}

The time evolution of an observable $Q(t)$ is computed from the time-dependent state $|\Psi(t)\rangle$
after an excitation induced by a small boost on the ground-state $|0\rangle$:
\begin{equation}
|\Psi(0)\rangle = e^{-i\epsilon \hat{Q}} |0\rangle=|0\rangle-i\epsilon\hat{Q}|0\rangle+O(\epsilon^2).
\label{eq:boost}
\end{equation}
Inserting $\hat{1}=\sum_\nu |\nu\rangle\langle\nu|$ where $\{|\nu\rangle\}$ are the eigenstates of $\hat{H}$ with eigenenergies $\{E_\nu\}$, 
$$|\Psi(0)\rangle =|0\rangle-i\epsilon\sum_\nu q_\nu|\nu\rangle+O(\epsilon^2).
$$

The time-evolution of the state reads
\begin{eqnarray}
|\Psi(t)\rangle &=& e^{-i\hat{H} t/\hbar}|\Psi(0)\rangle \nonumber \\
&=&e^{-iE_0t/\hbar} \left( |0\rangle - i\epsilon\sum_\nu q_\nu e^{-i\omega_\nu t} |\nu\rangle  \right) + O(\epsilon^2),\nonumber \\
&&
\end{eqnarray}
with $\hbar\omega_\nu = E_\nu-E_0$.
The response to this excitation can be written
\begin{eqnarray}
Q(t)&=&\langle\Psi(t)|\hat{Q}|\Psi(t)\rangle-\langle0|\hat{Q}|0\rangle \nonumber\\
&=& \left( \langle0| + i\epsilon\sum_\nu q_\nu^* e^{i\omega_\nu t} \langle\nu|  \right) 
\hat{Q}
\left( |0\rangle - i\epsilon\sum_\nu q_\nu e^{-i\omega_\nu t} |\nu\rangle  \right)
-\langle0|\hat{Q}|0\rangle +O(\epsilon^2) \nonumber\\
&=&-2\epsilon\sum_\nu |q_\nu|^2\sin \omega t+O(\epsilon^2).
\label{eq:linresp}
\end{eqnarray}

The energies $\hbar\omega_\nu$ and transition amplitudes $q_\nu$ can be extracted from the strength function
\begin{equation}
R_{Q}(\omega) =\lim_{\epsilon\rightarrow 0} \frac{-1 }{\pi \epsilon}\,
\int_{0}^{\infty}  dt\, {Q}(t) \, \sin (\omega t). \label{eq:strengthlin} 
\end{equation}
Indeed, using the expression (\ref{eq:linresp}) for $Q(t)$,
\begin{eqnarray}
R_{Q}(\omega) &=&  \frac{2}{\pi}\sum_\nu \, |q_\nu |^2 \int_0^\infty dt\, \sin(\omega t) \sin(\omega_\nu t)\\
&=& \sum_\nu \, |q_\nu |^2  \delta (\omega - \omega_\nu). \label{eq:strengthfinal}
\end{eqnarray}

\subsection{Strength function from TDHF evolution}

The evolution of $Q(t)$ can be computed directly from TDHF codes. 
The boost of Eq.~(\ref{eq:boost}) can be applied directly to the single-particle wave-functions according to 
$$
|\Psi(0)\rangle = e^{-i\epsilon \hat{Q}} |0\rangle=e^{-i\epsilon \sum_{i=1}^A\hat{q}(i)} |\varphi_1\cdots\varphi_A\rangle 
=|\left(e^{-i\epsilon {q}}\varphi_1\right)\cdots\left(e^{-i\epsilon {q}}\varphi_A\right)\rangle.
$$
To ensure that the response is in the linear regime, various values are usually used to check that $\langle\hat{Q}(t)\rangle\propto \varepsilon$.

\begin{figure}[htbp] 
   \centering
   \includegraphics[width=4.6in]{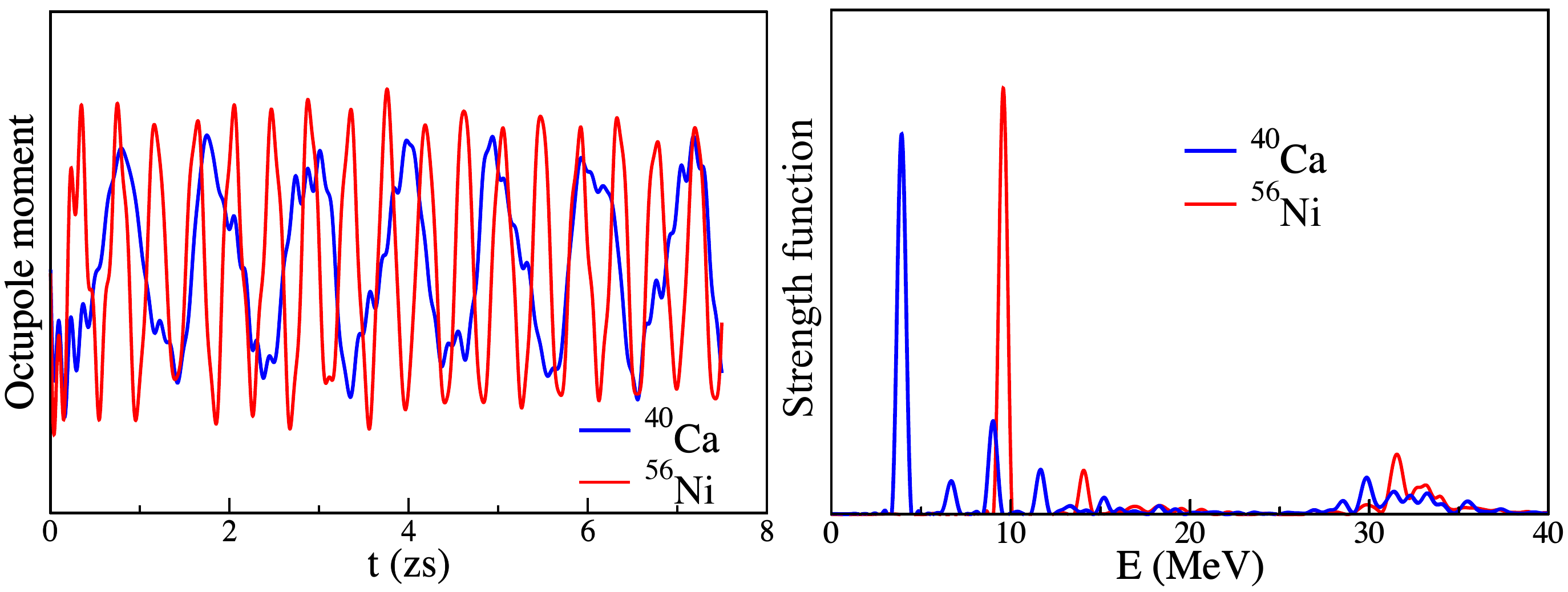} 
   \caption{(left) Time-evolution of the octupole moment following an octupole boost in $^{40}$Ca and $^{56}$Ni. (right) Corresponding strength functions. The vertical axes are in arbitrary units.}
   \label{fig:Octupole}
\end{figure}

An example of octupole response  $\langle\hat{Q}(t)\rangle$ in the linear regime is shown in Fig.~\ref{fig:Octupole}  in $^{40}$Ca and $^{56}$Ni.
The computation of the strength function can be done with Eq.~(\ref{eq:strengthlin}). 
However, the computational time being finite, the integral cannot be performed up to $t\rightarrow\infty$. 
Performing the integration up to $T$ without modifying Eq.~(\ref{eq:strengthlin}) would induce spurious oscillations in the spectrum as can be expected from the Fourier transform of a signal convoluted with a step function.  
To avoid such spurious oscillations, one usually multiply $Q(t)$ by a damping function $f(t)$ (e.g., a $\cos^2$ or a Gaussian function) decreasing slowly from $f(0)=1$ to $f(T)=0$. 
The damping function induces a small width proportional to $1/T$ in the strength function. 

The strength functions associated with the octupole responses in $^{40}$Ca and $^{56}$Ni are shown in Fig.~\ref{fig:Octupole}.
The rapid oscillation of the octupole moment in $^{56}$Ni is associated with a higher energy $\hbar\omega\sim9$~MeV than in $^{40}$Ca where the main low-lying octupole vibration is found at $\sim4$~MeV. Other states with smaller strengths are also observed at higher energy. 
As these nuclei have ground-state spin and parity $0^+$, and as the octupole operator allows for a change of angular momentum $\Delta L=3\hbar$, these vibrational states have a spin-parity $3^-$. 
The area of the peaks correspond to the  transition probabilities $|q_\nu|^2$.

\subsection{Reduced electric transition probability and deformation parameters}

The transition amplitudes can be used to evaluate other useful quantities such as the reduced electric transition probability and the deformation parameters.
The former can be evaluated according to 
$$B(E\lambda;0^+_{gs}\rightarrow \nu) =\frac{Z^2}{A^2}e^2|q_\nu|^2,$$
where the proton density is assumed to be proportional to the neutron density. 
For instance, one can evaluate $B(E2;0^+_{g.s.}\rightarrow 2^+)$ from the TDHF evolution of the quarupole moment $Q_{20}(t)$ following  a quadrupole boost. 

The deformation parameter can be computed from 
$$ \beta_\lambda^{(\nu)}=\frac{4\pi |q_\nu|}{3AR_0^\lambda},$$
where $R_0\simeq r_0A^{1/3}$ is the nuclear radius with $r_0\simeq1.1-1.2$~fm.
It is often useful to compare deformations of different nuclei, as well as to account for couplings to low-lying vibrations in coupled-channel calculations of heavy-ion collisions. 

\subsection{Giant resonances}

Giant resonances are highly collective nuclear vibrations that can usually decay by emitting nucleons. 
Their first phonons are often found between 10 and 30~MeV, depending on their multipolarity as well as their scalar/vector and isoscalar/isovector characteristics. 
Although most TDHF applications to nuclear vibrations are dedicated to giant resonances, the fact that they decay via particle emission means that such calculations has to be treated with care to avoid spurious finite size effects of the numerical box. 
To avoid using very large boxes (which often limit the calculations to spherical symmetric systems), one can use absorbing boundary conditions. 
In this case, the resulting strength functions correspond to continuum-RPA calculations.

\begin{figure}[htbp] 
   \centering
   \includegraphics[width=4.6in]{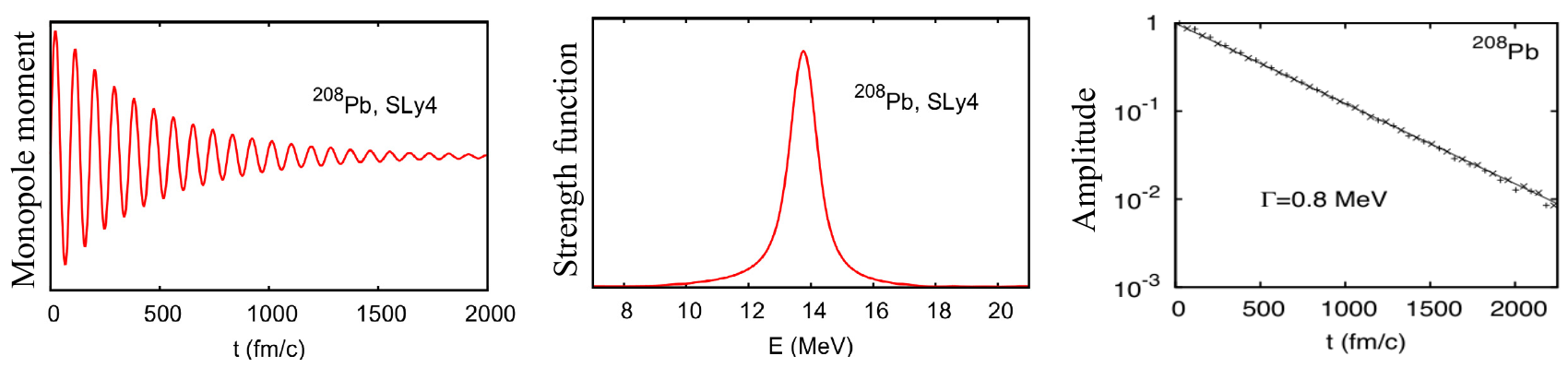} 
   \caption{(left) Time-evolution of the monopole moment following a monopole boost in $^{208}$Pb. (middle) Corresponding strength functions. (right) Amplitude of the oscillations (in logarithmic vertical scale). From (Avez 2009).}
   \label{fig:GRdecay}
\end{figure}

An example of TDHF calculation of a monopole giant resonance in $^{208}$Pb is shown in Fig.~\ref{fig:GRdecay}. 
The direct decay of the oscillation amplitude is due to evaporation of nucleon wave-functions that are leaving the nucleus. 
This decay is exponential and contributes to the width of the peak in the strength function. 

\subsection{RPA equation}

TDHF can be used in the
small amplitude regime to extract energies
and transition amplitudes of vibrational states
at the RPA level. This is not, however, how
a standard RPA code works.
The goal is now to show the connection between linearised TDHF and RPA in more details. 

When  a small boost $e^{-i\varepsilon\hat{Q}}$ is applied to  the
wave function, it induces a perturbation to the
one-body density matrix
\begin{equation}
\rho=\rho^{(0)}+\varepsilon\rho^{(1)}+O(\varepsilon^2) \label{eq:rho2}
\end{equation}
where 
$\rho^{(0)}=\sum_{h=1}^A |\varphi_h^{(HF)}\rangle\langle \varphi_h^{(HF)}|$
is the one-body density matrix of the Hartree-Fock ground-state. 
In TDHF, the many-body state remains a Slater with the corresponding 
one-body density matrix
$\rho=\sum_{i=1}^A |\varphi_i\rangle\langle \varphi_i|$.
One can see that $\rho$ is a projector onto the sub-space of occupied single-particle 
states by computing $\rho^2= \sum_{i,j=1}^A |\varphi_i\rangle\langle \varphi_i| \varphi_j\rangle\langle \varphi_j|
=\rho$ as $\langle \varphi_i| \varphi_j\rangle=\delta_{ij}$. 
Using this property with the expression~(\ref{eq:rho2}) gives
\begin{eqnarray}
\rho^2&=&{\rho^{(0)}}^2+\varepsilon\left(\rho^{(0)}\rho^{(1)}+\rho^{(1)}\rho^{(0)}\right)+O(\varepsilon^2) \nonumber\\
=\rho&=&\rho^{(0)}+\varepsilon\rho^{(1)}+O(\varepsilon^2).\nonumber
\end{eqnarray}
As a result,  ${\rho^{(0)}}^2=\rho^{(0)}$ and
\begin{equation}
\rho^{(1)}=\rho^{(0)}\rho^{(1)}+\rho^{(1)}\rho^{(0)}.\label{eq:rho1}
\end{equation}

In the single-particle basis which diagonalises $\rho^{(0)}$, 
$$\rho^{(0)}=\begin{pmatrix}
\left(\rho^{(0)}_{hh}\right) &\left(\rho^{(0)}_{hp}\right) \\
\left(\rho^{(0)}_{ph}\right) &\left(\rho^{(0)}_{pp}\right) \\
\end{pmatrix}
=\begin{pmatrix}
\mathbb{1} & \mathbb{0} \\
\mathbb{0} &\mathbb{0} \\
\end{pmatrix}$$
where $h$ and $p$ denote hole and particle states, respectively.

Let us write $\rho^{(1)}$ in this basis and use Eq.~(\ref{eq:rho1}) to get
$$\rho^{(1)}=\begin{pmatrix}
\left(\rho^{(1)}_{hh}\right) &\left(\rho^{(1)}_{hp}\right) \\
\left(\rho^{(1)}_{ph}\right) &\left(\rho^{(1)}_{pp}\right) \\
\end{pmatrix}
=\begin{pmatrix}
2\left(\rho^{(1)}_{hh}\right) &\left(\rho^{(1)}_{hp}\right) \\
\left(\rho^{(1)}_{ph}\right) &\mathbb{0} \\
\end{pmatrix},
$$
implying that $\rho^{(1)}_{hh}=\rho^{(1)}_{pp}=0$. 
As a result, $\rho^{(1)}$ has only particle-hole non-zero matrix elements. 

The RPA equation is an equation for $\rho^{(1)}$. It is obtained
from the linearisation of the TDHF equation, i.e., keeping only
terms linear in $\varepsilon$:
\begin{equation}
i\hbar\frac{d}{dt}{\rho}= \left[{h},{\rho}\right] \,\,\,\Rightarrow \,\,\,
i\hbar\frac{d}{dt}\left(\rho^{(0)}+\varepsilon\rho^{(1)}\right)= \left[h[\rho^{(0)}+\varepsilon\rho^{(1)}],\rho^{(0)}+\varepsilon\rho^{(1)}\right] .
\end{equation}
Expanding the Hamiltonian as 
$$h[\rho^{(0)}+\varepsilon\rho^{(1)}]=h[\rho^{(0)}]+\frac{\delta h}{\delta \rho}\varepsilon\rho^{(1)}+O(\varepsilon^2)$$
leads to 
$$i\hbar\frac{d}{dt}\rho^{(1)}= \left[h[\rho^{(0)}],\rho^{(1)}\right] +\left[ \frac{\delta h}{\delta \rho}\rho^{(1)},\rho^{(0)}\right].
$$
Note that $\frac{\delta h}{\delta \rho}\rho^{(1)}$ is a shorthand notation for 
$$\sum_{ph}\left[\left.\frac{\delta h}{\delta \rho_{ph}}\right|_{\rho=\rho^{(0)}}\, \rho^{(1)}_{ph}
+\left.\frac{\delta h}{\delta \rho_{hp}}\right|_{\rho=\rho^{(0)}}\, \rho^{(1)}_{hp}\right].
$$
One then gets the RPA equation
\begin{equation}
i\hbar\frac{d}{dt}{\rho^{(1)}}= \mathcal{M}\rho^{(1)}, 
\end{equation}
where
$$\mathcal{M}\,\,\cdot \,\,= \left[h[\rho^{(0)}],\,\,\cdot\,\,\right]+\left[\frac{\delta h}{\delta \rho}\,\,\cdot\,\,,\rho^{(0)}\right]$$
is the RPA matrix acting only on the particle-hole  $(ph)$ space and
$\frac{\delta h}{\delta \rho}$ is the RPA, or $ph$, residual interaction. 

\subsection{RPA modes}

Let us decompose the $ph$ density matrix as 
$$\rho^{(1)}(t)=\sum_\nu\rho_\nu^{(1)} e^{i\omega_\nu t} +H.c. ,$$
where $H.c.$ stands for ``Hermitian conjugate''. 
The RPA equation then becomes
\begin{equation}
i\hbar\sum_\nu \left( \rho_\nu^{(1)} i\omega_\nu e^{i\omega_\nu t} - {\rho_\nu^{(1)}}^\dagger i\omega_\nu e^{-i\omega_\nu t}\right)=
\mathcal{M}\sum_\nu \left( \rho_\nu^{(1)}  e^{i\omega_\nu t} +{\rho_\nu^{(1)}}^\dagger  e^{-i\omega_\nu t}\right).
\end{equation}
leading to 
\begin{equation}
-\hbar\omega_\nu\rho_\nu^{(1)} =
\mathcal{M}\rho_\nu^{(1)}  \,\,\,\, \mbox{ and } \,\,\,\,\,
\hbar\omega_\nu{\rho_\nu^{(1)}}^\dagger =
\mathcal{M}{\rho_\nu^{(1)} }^\dagger.
\end{equation}
This is an eigenvalue problem with the RPA modes $\nu$ of energy $\hbar \omega_\nu$. 

These energies $\hbar\omega_\nu$ are the same as the ones obtained from the linear response of TDHF,  corresponding to peaks in the strength functions such as the ones plotted in Fig.~\ref{fig:Octupole}. 
The task of an RPA code is then to diagonalise the RPA matrix in order to get these energies, and, from the corresponding eigenstates, the transition amplitudes. Unlike TDHF, no explicit time evolution is required. 

\begin{figure}[htbp] 
   \centering
   \includegraphics[width=4.6in]{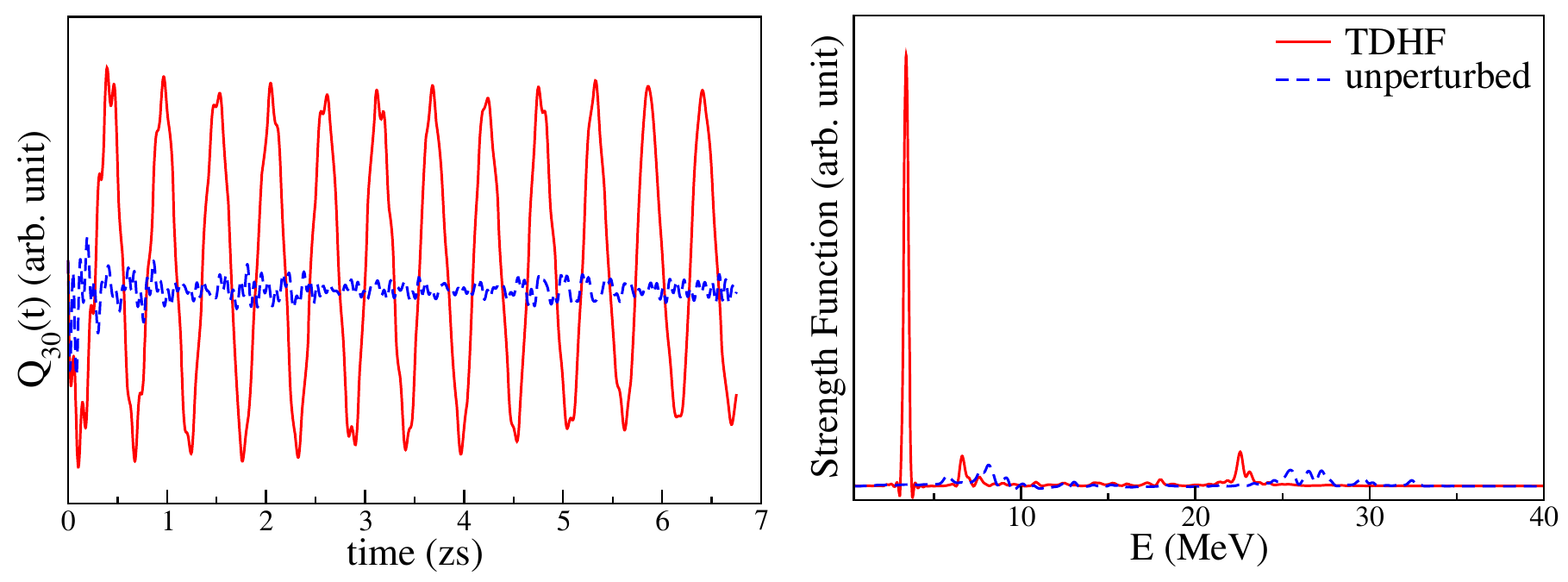} 
   \caption{(left) Time-evolution of the octupole moment following an octupole boost in $^{208}$Pb with (solid line) and without (dashed line) RPA residual interaction. (right) Corresponding strength functions. From (Simenel 2012).}
   \label{fig:residual}
\end{figure}

RPA goes beyond HF thanks to the residual interaction $\frac{\delta h}{\delta \rho}$. 
In HF, $[h, \rho] = 0$ which, in the basis which
diagonalises both $h$ and $\rho$, gives
$$h^{(0)}=\begin{pmatrix}
\left(h^{(0)}_{hh}\right) &\mathbb{0} \\
\mathbb{0} &\left(h^{(0)}_{pp}\right) \\
\end{pmatrix},
$$
with $\left(h^{(0)}_{hh}\right)$ and $\left(h^{(0)}_{pp}\right)$ diagonal matrices with single-particle energies $e_h$ and $e_p$ on their diagonal.
In RPA, there is also the residual interaction 
$$\frac{\delta h[\rho]_{ph}}{\delta \rho_{p'h'}}=\frac{\delta^2E[\rho]}{\delta\rho_{hp}\delta\rho_{p'h'}}$$
 in the particle-hole channel. 
This residual interaction is what is responsible for the collectivity of vibrational state, allowing many 1p1h excitations to contribute coherently. 
This is illustrated in Fig.~\ref{fig:residual} where a RPA response  (from TDHF) is compared to an ``unperturbed'' one obtained without the residual interaction. Large peaks in the strength function, corresponding to collective vibrations, are indeed only observed once the RPA residual interaction is taken into account.

 Collective excitations in RPA (and in TDHF in the linear regime) are then coherent superpositions of one-particle one-hole $(1p1h)$ excitations.
One  then understands why large differences between strength functions of various nuclei are sometimes found, as illustrated in the octupole responses of $^{40}$Ca and $^{56}$Ni in Fig.~\ref{fig:Octupole}. 
Indeed, despite the fact that these nuclei are both $N=Z$ doubly magic isotopes, their low-lying collective octupole vibrations have very different properties, with the $3^-_1$ energy in $^{56}$Ni more than twice the one in $^{40}$Ca.
This is because in $^{40}$Ca, $1p1h$ configurations coupling to $3^-$ spin-parity can be formed by promoting nucleons from the $sd$ shells to the empty $1f_{7/2}$ level sitting just above the Fermi level. 
In $^{56}$Ni, however, the $1f_{7/2}$ level is fully occupied, and coupling a hole in this level to a particle in the other $fp$ states cannot produce the desired spin-parity. 
As a result,  the first $3^-$ configurations can only be obtained by coupling $1f_{7/2}$ holes with the particle states in the higher energy $1g_{9/2}$ level, and/or by coupling $sd$ holes with $fp$ (excluding $1f_{7/2}$) particles states, producing a collective $3^-_1$ states at much higher energy than in $^{40}$Ca.

\subsection{Widths of giant resonances}

The width of a giant resonance (GR) has 3 components:
\begin{itemize}
\item The {\it Landau damping}s is due to the coupling of
the collective (coherent) superposition of $1p1h$ configurations forming the GR to incoherent $1p1h$ excitations.
\item The {\it escape width} is due to the emission of particles.
It requires a proper treatment of the continuum.
\item The {\it spreading width} comes from the coupling of $1p1h$
configurations  to $2p2h$ states. 
\end{itemize}

 RPA and TDHF account for the Landau damping and the escape
width (if the continuum is properly accounted for), but
not the spreading width which requires $2p2h$ residual
interaction. This is the purpose of the ``second-RPA'' and extended TDHF approaches. 
As a result, the RPA and TDHF are known to underestimate the width of GR.

\subsection{TDHF versus RPA}

Although TDHF, in the linear limit, is formally equivalent to RPA, there are pros and cons in their implementation, and, depending on the application, one may favour the use of one solver against the other. 
\begin{itemize}
\item RPA is sometimes solved with a simplified residual
interaction in the $ph$ channel (e.g., neglecting
spin-orbit or  Coulomb interaction). 
This can lead to spurious modes 
which need to be removed. TDHF, on the contrary,
gives a fully self-consistent RPA response, i.e., the
same interaction (or functional) is used to
compute $\rho^{(0)}$ and $\rho^{(1)}$.
Naturally, fully self-consistent RPA codes that account for all terms in the residual interaction are free of such spurious modes. 
\item Unlike RPA, TDHF is not limited to the linear
response. Nonlinearities in TDHF can be used
to investigate other terms of the residual interaction
than $1p1h$, such as $3p1h$ and $3h1p$.
\item A proper treatment of the continuum is necessary
for unbound states such as giant resonances.
In continuum-RPA, this is done with Coulomb
(for protons) and Hankel (for neutrons) functions.
In TDHF, this is often done with absorbing boundary
conditions.
\item RPA gives the amplitudes of the single-particles
contributing to a collective state more directly than
TDHF.
\item TDHF may require long computational times, in particular if one wants to achieve a high resolution in the strength functions.
\end{itemize}

\section{Time-Dependent Hartree-Fock-Bogoliubov theory}
\label{sec:TDHFB}

Pairs of nucleons with a strong overlap of their wave-functions (e.g., Cooper pairs of nucleons in time reversed states
$|j\,\,m\rangle$ and $|j\,\,-\!m\rangle$) are particularly sensitive to the short-range part of the nuclear interaction, leading to
a pairing energy and pairing correlations between
the particles: if one particle goes from one energy
level to another, then the other one is expected to
follow. The resulting correlated state is a coherent superposition of $2p2h$ configurations. 

\subsection{Manifestation of pairing}

In condensed matter, Cooper pairs are responsible for superconductivity and superfluidity.
In nuclear physics, pairing induces odd-even mass staggering
(even numbers of neutrons or protons are more bound
than odd numbers), enhanced pair transfer in
heavy-ion collisions (with respect to the transfer of 2
independent nucleons), backbending effect (moment
of inertia suddenly increases when there is enough rotational
energy to break a pair, which in turns slows down
the rotation), and a first (non-collective) excitation in mid-shell
even-even nuclei at $\sim2$~MeV (energy needed to break
a pair).

Note that the pairing residual interaction can only move
a pair by $\sim2$~MeV, so only nucleons near the Fermi surface can be paired.
Indeed, the scattering of more bound nucleons would be blocked 
by the Pauli exclusion principle as the final state would be already occupied.  
As a result, the contribution of pairing to the total binding
energy is relatively small. 

\subsection{Including correlations via symmetry breaking}

Before discussing the specific case of pairing, let us review the general technique of including correlations at the mean-field level through symmetry breaking. 
The idea is that by allowing the mean-field Hamiltonian to break a symmetry of the exact Hamiltonian, one can include some of the residual interaction, while preserving the simplicity of a mean-field treatment. 

\subsubsection{Translational invariance}

As a first example, let us discuss the case of  translational invariance. 
The interaction between two nucleons depends on their
relative distance, not the position of their centre of mass,
so the interaction is of the form $V(|\mathbf{r}_1-\mathbf{r}_2|)$.
If translational invariance is imposed to the mean-field
Hamiltonian 
$\hat{H}_{MF}=\sum_{i=1}^A\hat{h}[\rho](i)$, however, one can see that
$\hat{h}[\rho]$, and therefore $\rho$ itself have to be flat (constant in space). 
In this case
the eigenstate of $\hat{h}$ are plane waves occupying the
entire space: There is no spatial correlations
between the nucleons and therefore no nucleus...

To allow a mean-field treatment of the nucleus,
 translational invariance needs then to be broken with
a position dependent mean-field potential trapping
the nucleons.

\subsubsection{Rotational invariance}

The same approach can be used by breaking rotational
invariance of $\hat{H}_{MF}$, i.e., with a deformed
mean-field along a particular direction.
This induces a deformation of
$\rho$ and allows for a mean-field
treatment of long-range correlations
responsible for static deformations
in nuclei.

\subsection{Pairing correlations through breaking of gauge invariance}

A mean-field
treatment of pairing correlations can be obtained by breaking
gauge invariance responsible for particle number
conservation.

\subsubsection{Generalised one-body density matrix}

Recall that the one-body density $\rho$ associated with a
Slater $|\Phi\rangle=\hat{a}_A^\dagger\cdots\hat{a}_1^\dagger |-\rangle$ 
contains the same information
as the Slater itself. One can then focus the reasoning on
how to treat pairing with an object like $\rho$.

A Slater has a ``good'' (i.e., a well defined) number of particles $A$. 
Thus, only the matrix elements $\rho_{\alpha\beta}=\langle\Phi|\hat{a}_\beta^\dagger\hat{a}_\alpha|\Phi\rangle$ are needed 
as $\langle\Phi|\hat{a}_\beta^\dagger\hat{a}_\alpha^\dagger|\Phi\rangle=\langle\Phi|\hat{a}_\beta\hat{a}_\alpha|\Phi\rangle=0$.
If one now considers a state which does not have a good number of particles, then the so-called {\it anomalous density}
$\kappa_{\alpha\beta}=\langle\Phi|\hat{a}_\beta\hat{a}_\alpha|\Phi\rangle$ and its complex conjugate
 $\kappa_{\alpha\beta}^*=\langle\Phi|\hat{a}_\alpha^\dagger\hat{a}_\beta^\dagger|\Phi\rangle$
do not vanish. In this case, a ``generalised'' one-body density matrix $\mathcal{R}$
which contains both $\rho$ and $\kappa$ is used.
It is defined as 
$$\mathcal{R}=\begin{pmatrix}
\rho&\kappa \\
-\kappa^* &1-\rho^* \\
\end{pmatrix}.$$

\subsubsection{Quasi-particle vacuum}

The focus is now on the state $|\Psi\rangle$. For a Slater, $|\Phi\rangle=\hat{a}_A^\dagger\cdots\hat{a}_1^\dagger |-\rangle$. 
In order to preserve the simplicity of mean-field equations, it is desirable to keep a
similar form of a product of independent operators. 
However, to describe a system that does not have a good number of particles,  a 
Slater cannot be used.

The form of a state describing a coherent superposition of various numbers
of pairs of particles can be guessed as something like
\begin{equation}
|\Psi\rangle\propto|-\rangle+\sum_{ij}(V_{ij}\hat{a}_i^\dagger\hat{a}_j^\dagger)|-\rangle+
\sum_{ijkl}(V_{ij}\hat{a}_i^\dagger\hat{a}_j^\dagger)(V_{kl}\hat{a}_k^\dagger\hat{a}_l^\dagger)|-\rangle+\cdots
\end{equation}
where the first term in the r.h.s. contains zero pair, the second contains one pair, the third contains two pairs, etc. 

This state can also be written
\begin{equation}
|\Psi\rangle\propto\Pi_{ij}(U_{ij}+V_{ij}\hat{a}_i^\dagger\hat{a}_j^\dagger)|-\rangle \label{eq:U+Vaa}
\end{equation}
or, equivalently, 
\begin{equation}
|\Psi\rangle\propto\Pi_{j}\left(\sum_iU_{ij}^*\hat{a}_i+V_{ij}^*\hat{a}_i^\dagger\right)|-\rangle \label{eq:Ua+Va}
\end{equation}
Note that, $U$ and $V$ are generic matrices. Their complex conjugation in Eq.~(\ref{eq:Ua+Va}) is a convention.
To show that Eqs.~(\ref{eq:U+Vaa}) and~(\ref{eq:Ua+Va}) are equivalent, start from (\ref{eq:Ua+Va}), develop, bring the $\hat{a}^\dagger$ to the left using $\{\hat{a}^\dagger_i,\hat{a}^\dagger_j\}=\{\hat{a}_i,\hat{a}_j\}=0$ and $\{\hat{a}^\dagger_i,\hat{a}_j\}=\delta_{ij}$, and use $\hat{a}|-\rangle=0$. 
This is easier to show at the BCS
level where $U$ and $V$ couple time reversed states $i$ and $\bar{i}$
only:
\begin{eqnarray}
|\Psi\rangle&\propto&\Pi_i \left(u_i \hat{a}_i-v_i\hat{a}^\dagger_{\bar{i}}\right)\left(u_i \hat{a}_{\bar{i}}+v_i\hat{a}^\dagger_i\right)|-\rangle\nonumber\\
&\propto&\Pi_i \left(u_i^2 \hat{a}_i\hat{a}_{\bar{i}}+u_iv_i \hat{a}_i\hat{a}^\dagger_i-v_iu_i \hat{a}^\dagger_{\bar{i}}\hat{a}_{\bar{i}}-v_i^2\hat{a}^\dagger_{\bar{i}}\hat{a}^\dagger_i\right)|-\rangle\nonumber\\
&\propto&\Pi_i \left(u_i+v_i\hat{a}^\dagger_i\hat{a}^\dagger_{\bar{i}}\right)|-\rangle.\nonumber
\end{eqnarray}

Introducing the quasiparticle operator $\hat{\beta}_i=\sum_i\left(U_{ij}^*\hat{a}_i+V_{ij}^*\hat{a}_i^\dagger\right)$ through a Bogoliubov transformation, one gets $|\Psi\rangle=\Pi_i\hat{\beta}_i|-\rangle$. 
The $U$ and $V$ matrices can be chosen such that
the quasi-particle operators obey the Fermionic anticommutation
relationship $\{\hat{\beta}^\dagger_i,\hat{\beta}^\dagger_j\}=\{\hat{\beta}_i,\hat{\beta}_j\}=0$ and $\{\hat{\beta}^\dagger_i,\hat{\beta}_j\}=\delta_{ij}$.
In this case,  $\hat{\beta}_i|\Psi\rangle=0$, $\forall i$, i.e., $|\Psi\rangle$ is a quasiparticle vacuum. 
A state with, say, N
quasiparticle excitations, $\beta_1^\dagger\cdots\beta_N^\dagger|\Psi\rangle$, is clearly a
state of independent quasi-particles. This is the generalisation of a Slater that  was looked for.

\subsubsection{Non-conservation of particle number}

Before going to the TDHFB equation, let us discuss why 
the particle number conservation had to be broken in the first place. 
The problem comes from the $2p2h$ component of $|\Psi\rangle$
which is associated with the two-body density matrix with
elements $\langle\hat{a}^\dagger_p\hat{a}^\dagger_{p'}\hat{a}_h\hat{a}_{h'}\rangle$. 
In (TD)HFB, these terms are
approximated by $\langle\hat{a}^\dagger_p\hat{a}^\dagger_{p'}\rangle\langle\hat{a}_h\hat{a}_{h'}\rangle$ (involving $\kappa$ and $\kappa^*$) which, to be non-zero,
require the state to be a superposition of Slaters with
different particle numbers, and which is described as
a quasi-particle vacuum.

\subsubsection{TDHFB equation}

The TDHFB equation can be obtained from the variational
principle $\delta S=0$ with the Dirac action and the
variational space defined by independent quasiparticles. Note,
however, that this variational space span subspaces of
$\mathcal{H}_N, \mathcal{H}_{N\pm2}, \mathcal{H}_{N\pm4}, \cdots$.  

The TDHFB equation is given here without derivation:
\begin{equation}
i\hbar\frac{d}{dt}\mathcal{R}=\left[\mathcal{H},\mathcal{R}\right],
\end{equation}
where $\mathcal{H}[\mathcal{R}]$ is the self-consistent generalised Hamiltonian of the form 
$$\mathcal{H}=\begin{pmatrix}
h&\Delta \\
-\Delta^* &-h^* \\
\end{pmatrix},$$
and 
$$\Delta_{ij}=\frac{\delta E[\rho,\kappa,\kappa^*]}{\delta\kappa_{ij}}$$
is the pairing field. 

\subsubsection{Static HFB equation}

It  can be shown that $\langle\hat{N}\rangle$ is in fact conserved in TDHFB, 
i.e., the particle number is conserved in average. 
However this is not the case in the algorithms used to solve the static HFB equation
\begin{equation}
\left[\mathcal{H},\mathcal{R}\right]=0.
\end{equation}
It is therefore necessary to add a
constraint on the particle number, e.g., replacing $\hat{H}$
by $\hat{H}+\lambda(\hat{N}-N_0)^2$, where $\lambda$ is a Lagrange parameter adjusted 
to force the system to have
$\langle\hat{N}\rangle= N_0$ particles at its minimum of energy.

\subsection{Restoring a good particle number}

Physicists are not always at ease with extracting
observables (wether for structure or reaction studies)
from states which have a number of particles
defined only in average. Therefore they sometimes
use a particle number projection technique to transform
the HFB state into a state with well defined
particle number.
The projector onto a state with $N$ particles acts as 
$$\hat{P}_N|\Psi\rangle=\frac{1}{2\pi}\int_0^{2\pi}d\varphi \,e^{i\varphi(\hat{N}-N)}|\Psi\rangle.$$
Indeed, decomposing $|\Psi\rangle=\sum_n C_n |\Psi_n\rangle$ with $\hat{N}|\Psi_n\rangle=n|\Psi_n\rangle$ and $n\in\mathbb{N}$,
one gets
$$\hat{P}_N|\Psi\rangle=\frac{1}{2\pi}\sum_n C_n\int_0^{2\pi}d\varphi \,e^{i\varphi(n-N)}|\Psi_n\rangle.$$
Noting that $\int_0^{2\pi}d\varphi \,e^{i\varphi(n-N)}=0$ for $n\ne N$, one finally obtains
$$\hat{P}_N|\Psi\rangle=C_N|\Psi_N\rangle.$$

A potential problem is that $\int_0^{2\pi}d\varphi \,e^{i\varphi\hat{N}}|\Psi\rangle$ is a sum over quasiparticle vaccua.
This takes the state outside
the variational space of independent quasiparticle states.
As a result, the final projected state is not entirely
consistent with the variational approach.

\subsection{Some applications of TDHFB}

\subsubsection{pairing vibrations} 
Pairing vibrations are associated with oscillations of
$\kappa(t)$. They can be studied in the same way as
``normal'' vibrations with $\rho(t)$ using the small amplitude
limit. The quasiparticle RPA (QRPA) can also be 
obtained from the linearisation of the TDHFB equation. 
Pairing vibrations are probed in pair transfer reactions
with an operator containing  $\hat{a}^\dagger\hat{a}^\dagger$ and/or $\hat{a}\hat{a}$ terms.

For instance, 
$$\hat{F}=\int d^3r\,f(r) \left(\hat{a}^\dagger(\mathbf{r},\uparrow)\hat{a}^\dagger(\mathbf{r},\downarrow)+\hat{a}(\mathbf{r},\uparrow)\hat{a}(\mathbf{r},\downarrow)\right)$$
induces $\Delta L=0$ transitions, e.g., from $0^+$
ground-state  of a
nucleus with $A$ nucleons to $0^+$ states in nuclei
with $A\pm2$ nucleons. The TDHFB linear response $\langle \hat{F}(t)\rangle$ contains
modes from both final nuclei. As in RPA, some modes
are clearly collective (larger peaks in the strength function).
They are called pairing vibrations and are expected to be
significantly populated in pair-transfer reactions.

\subsubsection{Fusion barrier}
Recently, full TDHFB simulations of fusion reactions have shown
a possible hindrance of fusion due to different gauge
angles (created by the gauge symmetry breaking) between
the colliding nuclei and producing a domain wall
at the neck (Magierski 2017; Scamps 2018).

\section{Balian-V\'en\'eroni variational principle}
\label{sec:BV}

The possibility to have different variational
principles was  mentioned earlier. Indeed the action is not unique and, as
long as one recovers Schr\"odinger in the exact case,
i.e., without restriction of the variational space, they could
equivalently be considered in order to build approximated dynamics.
It turns out that TDHF is usually recovered at the
simplest level, both from variational and non-variational
approaches. It is often interesting to see different approaches
to derive an equation such as TDHF as they usually give us
new perspectives on the range of possible applications and
limitations of the theory.

\subsection{Balian-V\'en\'eroni action}

The Dirac action has two variational quantities, $\Psi(t)$
and $\Psi^*(t)$, both describing the time evolution
of the many-body state. They lead to the Schr\"odinger
 picture in which the time evolution is carried by
the state while the observables are time-independent quantities. 
This is not the only possible
picture. For instance, in the Heisenberg picture, the
observables evolve in time while the state remains
static.

Balian and V\'en\'eroni (BV) proposed
an action that mixes both Schr\"odinger and Heisenberg pictures.
The BV action has two variational quantities, namely the
many-body state described by its density matrix
$$\hat{D}(t)=|\Psi(t)\rangle\langle\Psi(t)|
,$$ 
and the observable $\hat{A}(t)$ (Balian and V\'en\'eroni 1981). This
way, one can recover the exact dynamics in both the
Schr\"odinger and Heisenberg pictures. The BV action
is defined as
\begin{equation}
S = \mathrm{Tr}\left[ \hat{A}(t_1)\hat{D}(t_1) \right]- \int_{t_0}^{t_1}  dt \,\mathrm{Tr}\!\left[ \hat{A}(t) \left(\frac{d\hat{D}(t)}{dt} + \frac{i}{\hbar}[\hat{H},\hat{D}(t)]\right) \right]
\label{eq:JA}
\end{equation}
with the boundary conditions 
\begin{equation}
\hat{D}(t_0)=\hat{D}_0,
\label{eq:BCD}
\end{equation}
(the initial state of the system $\hat{D}_0$ at the initial time $t_0$ is known) and
\begin{equation}
\hat{A}(t_1) = \hat{A}_1,
\label{eq:BCA}
\end{equation}
(the goal is to compute the expectation value $\langle\hat{A}_1\rangle$ at the final time $t_1>t_0$). 

\subsection{Exact evolution}

\subsubsection{Schr\"odinger equation}

Let us verify that  the exact evolution is recovered if no restrictions on  the variational spaces are imposed.
As the approach contains two variational quantities, the variational principle $\delta S=0$ is obtained by
 imposing $\delta_A S=0$ and
 $\delta_D S=0$, where $\delta_X$ induces small variations with respect to $\hat{X}(t)$.
 
 As $\hat{A}(t_1)$ is fixed by a boundary condition, $\delta_A\,\left[\hat{A}(t_1)\hat{D}(t_1) \right]=0$. 
 As a result, 
$$ \int_{t_0}^{t_1}  dt \,\mathrm{Tr}\!\left[ \delta_A\hat{A}(t) \left(\frac{d\hat{D}(t)}{dt} + \frac{i}{\hbar}[\hat{H},\hat{D}(t)]\right) \right]=0.$$

For this to be true for all variations of $\hat{A}(t)$, the term in brackets must be zero, 
leading to the Liouville form of the Schr\"odinger equation
\begin{equation}
i\hbar\frac{d\hat{D}(t)}{dt}=\left[\hat{H},\hat{D}(t)\right].
\label{eq:Liouville}
\end{equation}

\subsubsection{Heisenberg equation}

Now consider the variations with respect to $\hat{D}(t)$.
It is easier to first rewrite the integral term in the  action~(\ref{eq:JA}) using an integration by part:
\begin{eqnarray}
\int_{t_0}^{t_1}  dt \,\mathrm{Tr}\!\left[ \hat{A}(t) \left(\frac{d\hat{D}(t)}{dt} + \frac{i}{\hbar}[\hat{H},\hat{D}(t)]\right) \right]  
=\mathrm{Tr}\left[ \hat{A}(t_1)\hat{D}(t_1)\right]-\mathrm{Tr}\left[ \hat{A}(t_0)\hat{D}(t_0) \right]&&\nonumber\\
- \int_{t_0}^{t_1} dt\, \mathrm{Tr}\left[ \hat{D}(t) \frac{d\hat{A}(t)}{dt}  \right]
+\frac{i}{\hbar} \int_{t_0}^{t_1} dt\, \mathrm{Tr}\left(\hat{A}(t) [\hat{H},\hat{D}(t)]\right).&&
\end{eqnarray}
As $\hat{D}(t_0)$ is fixed, the variation of $\mathrm{Tr}\left[ \hat{A}(t_0)\hat{D}(t_0) \right]$ vanishes.
Conveniently, the $\mathrm{Tr}\left[ \hat{A}(t_1)\hat{D}(t_1)\right]$ will cancel with the first term in the action (\ref{eq:JA}). 
In addition, the last term can be rearranged using 
 $$\mathrm{Tr}(\hat{A}[\hat{H},\hat{D}])=-\mathrm{Tr}(\hat{D}[\hat{H},\hat{A}]).$$
 One then obtains
 \begin{eqnarray}
\delta_DS&=& \int_{t_0}^{t_1} dt\, \mathrm{Tr}\left[ \delta_D\hat{D}(t) \left(\frac{d\hat{A}(t)}{dt} + \frac{i}{\hbar}[\hat{H}(t),\hat{A}(t)]\right) \right]=0.
\end{eqnarray}
As this must be true for all variation of $\hat{D}(t)$, one concludes that
\begin{equation}
i\hbar\frac{d\hat{A}(t)}{dt}=\left[\hat{H},\hat{A}(t)\right],
\label{eq:Ehrenfest}
\end{equation}
which gives the exact evolution of $\hat{A}(t)$ in the Heisenberg picture. 

\subsection{TDHF from the BV variational principle}

Let us restrict the variational space of $\hat{D}(t)$ to independent particle states, i.e., $\hat{D}=|\Phi\rangle\langle\Phi|$ where $|\Phi\rangle$ is a Slater, and the variational space of $\hat{A}(t)$ to one-body operators, i.e., in second quantisation, 
$$\hat{A}=\sum_{\alpha\beta}A_{\alpha\beta}\hat{a}^\dagger_\alpha\hat{a}_\beta.$$

Because the variations are arbitrary, one is  free
to choose 
$$\delta_A\hat{A}(t)=\hat{a}^\dagger_\alpha\hat{a}_\beta, \,\,\,\,t_0\le t\le t_1,$$
where $\alpha$ and $\beta$ are arbitrary.

$\delta S_A=0$ leads to 
\begin{eqnarray}
\mathrm{Tr}\left[\hat{a}^\dagger_\alpha\hat{a}_\beta\left(\frac{d}{dt}|\Phi\rangle\langle\Phi|+\frac{i}{\hbar}[\hat{H},|\Phi\rangle\langle\Phi|]\right)\right]&=&0.\nonumber\\
\frac{d}{dt}\mathrm{Tr}\left(\hat{a}^\dagger_\alpha\hat{a}_\beta|\Phi\rangle\langle\Phi|\right)
+\frac{i}{\hbar}\mathrm{Tr}\left(\hat{a}^\dagger_\alpha\hat{a}_\beta\left[\hat{H},|\Phi\rangle\langle\Phi|\right]\right)&=&0.\label{eq:traces}
\end{eqnarray}
Form the definition of the trace $\mathrm{Tr}(\hat{O})=\sum_\nu\langle\Psi_\nu|\hat{O}|\Psi_\nu\rangle$, where $\{|\Psi_\nu\rangle\}$ is a basis of $\mathcal{H}_N$,  one has
\begin{equation}\mathrm{Tr}\left(\hat{a}^\dagger_\alpha\hat{a}_\beta|\Phi\rangle\langle\Phi|\right)=
\sum_\nu\langle\Psi_\nu|\hat{a}^\dagger_\alpha\hat{a}_\beta|\Phi\rangle C_\nu^*=\langle\Phi|\hat{a}^\dagger_\alpha\hat{a}_\beta|\Phi\rangle=\rho_{\beta\alpha},\label{eq:Trrho}
\end{equation}
 where the state is decomposed into $|\Phi\rangle=\sum_\nu C_\nu|\Psi_\nu\rangle$ and the definition of the one-body density matrix in Eq.~(\ref{eq:defrho}) was used.

The second trace in Eq.~(\ref{eq:traces}) can be computed in the same way:
\begin{eqnarray}
\mathrm{Tr}\left(\hat{a}^\dagger_\alpha\hat{a}_\beta\left[\hat{H},|\Phi\rangle\langle\Phi|\right]\right)&=&
\mathrm{Tr}\left(\hat{a}^\dagger_\alpha\hat{a}_\beta\hat{H}|\Phi\rangle\langle\Phi|\right)-
\mathrm{Tr}\left(\hat{a}^\dagger_\alpha\hat{a}_\beta|\Phi\rangle\langle\Phi|\hat{H}\right)\nonumber\\
&=&\langle\Phi|\hat{a}^\dagger_\alpha\hat{a}_\beta\hat{H}|\Phi\rangle-
\mathrm{Tr}\left(\hat{H}\hat{a}^\dagger_\alpha\hat{a}_\beta|\Phi\rangle\langle\Phi|\right)\nonumber\\
&=&\langle\Phi|\hat{a}^\dagger_\alpha\hat{a}_\beta\hat{H}|\Phi\rangle-
\langle\Phi|\hat{H}\hat{a}^\dagger_\alpha\hat{a}_\beta|\Phi\rangle\nonumber\\
&=&\langle\Phi|\left[\hat{a}^\dagger_\alpha\hat{a}_\beta,\hat{H}\right]|\Phi\rangle.\label{eq:TrH}
\end{eqnarray}

The goal is now to show that this last term is just $[h,\rho]$.
This is usually done by introducing an explicit expression for $\hat{H}$. 
However, here, one wishes to remain general so that the approach is still valid in the EDF formalism.
 
Introducing a basis $\{|\nu\rangle\}$ of eigenstates of $\hat{H}$ with eigenvalues $\{E_\nu\}$, 
\begin{eqnarray}
\langle\Phi|\left[\hat{a}^\dagger_\alpha\hat{a}_\beta,\hat{H}\right]|\Phi\rangle&=&
\langle\Phi|\left[\hat{a}^\dagger_\alpha\hat{a}_\beta,\sum_\nu|\nu\rangle\langle\nu|\hat{H}\right]|\Phi\rangle\nonumber\\
&=&\sum_\nu E_\nu\langle\Phi|\left[\hat{a}^\dagger_\alpha\hat{a}_\beta,|\nu\rangle\langle\nu|\right]|\Phi\rangle\nonumber
\end{eqnarray}

One can write $|\nu\rangle$ in the $n-$particle $n-$hole basis built from $|\Phi\rangle$ as
\begin{equation}
|\nu\rangle=C_0^\nu|\Phi\rangle+\sum_{ph}C_{ph}^\nu \hat{a}_p^\dagger\hat{a}_h|\Phi\rangle
+\sum_{pp'hh'}C_{pp'hh'}^\nu \hat{a}_p^\dagger\hat{a}_{p'}^\dagger\hat{a}_h\hat{a}_{h'}|\Phi\rangle+\cdots \label{eq:nudecomp}
\end{equation}
with $C_{ph}^\nu=\langle\Phi|\hat{a}^\dagger_h\hat{a}_p|\nu\rangle$. 
As a result, 
\begin{equation}
\langle\Phi|\left[\hat{a}^\dagger_\alpha\hat{a}_\beta,\hat{H}\right]|\Phi\rangle
=\sum_\nu E_\nu\left(C_{\beta\alpha}^\nu {C_0^\nu}^*-C_0^\nu {C_{\alpha\beta}^\nu}^*\right).\label{eq:commoad}
\end{equation}

Now rearrange the $[h,\rho]$ term in order to express it as a function of the same quantities:
\begin{equation}
[h,\rho]_{\alpha\beta}=\sum_\gamma \left(h_{\alpha\gamma}\rho_{\gamma\beta}-\rho_{\alpha\gamma}h_{\gamma\beta}\right)
=\sum_\gamma \left(\frac{\delta E[\rho]}{\delta\rho_{\gamma\alpha}}\rho_{\gamma\beta}-\rho_{\alpha\gamma}\frac{\delta E[\rho]}{\delta\rho_{\beta\gamma}}\right)\nonumber
\end{equation}
Noting  $E[\rho]=\langle\Phi|\hat{H}|\Phi\rangle\equiv\langle\hat{H}\rangle$ and $\rho_{\alpha\beta}=\langle\Phi|\hat{a}_\beta^\dagger\hat{a}_\alpha|\Phi\rangle
\equiv\langle\hat{a}_\beta^\dagger\hat{a}_\alpha\rangle$, 
\begin{eqnarray}
\left[h,\rho\right]_{\alpha\beta} 
&=& \sum_\gamma \left(\frac{\delta \langle\hat{H}\rangle}{\delta\langle\hat{a}^\dagger_\alpha\hat{a}_\gamma\rangle}\langle\hat{a}^\dagger_\beta\hat{a}_\gamma\rangle - \langle\hat{a}^\dagger_\gamma\hat{a}_\alpha\rangle\frac{\delta \langle\hat{H}\rangle}{\delta\langle\hat{a}^\dagger_\gamma\hat{a}_\beta\rangle}\right) \label{eq:comm}.
\end{eqnarray}

Notice that, in the canonical basis, $\langle\hat{a}^\dagger_\beta\hat{a}_\gamma\rangle$ and $ \langle\hat{a}^\dagger_\gamma\hat{a}_\alpha\rangle$ are non-zero only if $\gamma$ is a hole state.  
Using $\langle\hat{H}\rangle=\sum_\nu E_\nu \langle\phi|\nu\rangle\langle\nu|\phi\rangle$ and the decomposition~(\ref{eq:nudecomp}), one can see  that Eq.~(\ref{eq:comm}) includes terms like
\begin{eqnarray}
\frac{\delta ( \langle\phi|\nu\rangle\langle\nu|\phi\rangle)}{\delta\langle\hat{a}^\dagger_\alpha\hat{a}_\gamma\rangle} &=& \sum_{ph} C^\nu_{ph} \frac{\delta  \langle\hat{a}^\dagger_p\hat{a}_h\rangle}{\delta\langle\hat{a}^\dagger_\alpha\hat{a}_\gamma\rangle} \langle\nu|\phi\rangle + \sum_{ph} {C^\nu_{ph}}^* \langle\phi|\nu\rangle \frac{\delta  \langle\hat{a}^\dagger_h\hat{a}_p\rangle}{\delta\langle\hat{a}^\dagger_\alpha\hat{a}_\gamma\rangle} \nonumber\\
&=& C^\nu_{\alpha\gamma}\langle\nu|\phi\rangle + {C^\nu_{\gamma\alpha}}^* \langle\phi|\nu\rangle. 
\end{eqnarray}
Only $C_{ph}$ terms are non zero and, for a Slater,  $\rho_{hh'}=\delta_{hh'}$ and $\rho_{\alpha p}=\rho_{p\alpha}=0$. 
As a result, $C^\nu_{\gamma\alpha}\rho_{\gamma\beta}=C^\nu_{\gamma\alpha}\rho_{\beta\gamma}=0$ 
as $\gamma$ cannot be a particle and a hole at the same time. 
Equation~(\ref{eq:comm}) then becomes
\begin{eqnarray}
\left[h,\rho\right]_{\alpha\beta}&=&\sum_\nu E_\nu \sum_\gamma \left(C_{\alpha{\gamma}}^\nu \langle\nu|\phi\rangle\rho_{\gamma\beta}-\rho_{\alpha{\gamma}} {C_{\beta{\gamma}}^\nu}^*\langle\phi|\nu\rangle\right)\nonumber\\
&=& \sum_\nu E_\nu \left(C_{\alpha{\beta}}^\nu {C_0^\nu}^*- {C_{\beta{\alpha}}^\nu}^*C_0^\nu\right),
\end{eqnarray}

Comparing with Eq.~(\ref{eq:commoad}), the conclusion is that 
\begin{equation}
\left[h,\rho\right]_{\alpha\beta}=\langle\phi|\left[\hat{a}^\dagger_\beta\hat{a}_\alpha,\hat{H}\right]|\phi\rangle.\label{eq:hrho}
\end{equation}
Combining Eqs.~(\ref{eq:traces}), (\ref{eq:Trrho}), (\ref{eq:TrH}), and (\ref{eq:hrho}), 
\begin{eqnarray}
\frac{d}{dt}\rho_{\beta\alpha}
+\frac{i}{\hbar}\left[h,\rho\right]_{\beta\alpha}&=&0\nonumber
\end{eqnarray}
which, after rearranging, leads to the TDHF equation in the Liouville form
\begin{equation}
i\hbar\frac{d}{dt}\rho=\left[h,\rho\right].\nonumber
\end{equation}

This demonstration shows that the BV
variational principle with independent particles and
one-body operator variational spaces leads to the TDHF
equation. Note that no explicit form for $\hat{H}$ was invoked, 
so this derivation still holds in 
 the  energy density functional approach. 

Importantly, the fact that TDHF was obtained by restricting the variational space for the observables to one-body operators tells us that TDHF
is only optimised to compute expectation values of one-body
observables.
In particular, fluctuations of one-body observables, 
$$\sigma_F=\sqrt{\langle\hat{F}^2\rangle-\langle\hat{F}\rangle^2},$$
 with $\hat{F}=\sum_{\alpha\beta}f_{\alpha\beta}\hat{a}_\beta^\dagger\hat{a}_\alpha$
 are outside the variational space.
Indeed, $\hat{F}^2$ contains two-body terms of the form  $\hat{a}^\dagger\hat{a}^\dagger\hat{a}\hat{a}$.
This explains why TDHF underestimates
the widths of fragment mass and charge distributions
in deep-inelastic collisions.

\subsection{Time-Dependent Random Phase Approximation}

To estimate one-body fluctuations and correlations, Balian and V\'en\'eroni
 used their variational principle with a larger
variational space for the operators $\hat{A}(t)$ (Balian and V\'en\'eroni 1984).
Instead of $\hat{a}^\dagger\hat{a}$ (one-body), they used a variational space for operators of the form $e^{\hat{a}^\dagger\hat{a}}$.
Then the fluctuations of an operator $\hat{F}$ can be computed from
$$\hat{A}_1=e^{-\varepsilon \hat{F}}$$
using
$$\ln \langle\hat{A}_1\rangle=-\varepsilon \hat{F}+\frac{1}{2}\varepsilon^2\left(\langle\hat{F}^2\rangle-\langle\hat{F}\rangle^2\right)+O(\varepsilon^3).$$
Indeed, the $\varepsilon^2$ term is $\sigma_F^2/2$.

How to solve the BV variational principle in this space
is beyond the scope of this chapter and can be found in details in (Simenel 2012). The
idea is to take small values of $\varepsilon$, which means
 small fluctuations of the action around
the stationary path are considered. By analogy with the RPA
(small fluctuation around a local static state minimising the energy), this approach is called the
time-dependent RPA (TDRPA). Note that it can be
also used to get correlations (in addition to fluctuations) between one-body
observables.

Correlations of two (commuting) one-body operators, $\hat{X}$ and $\hat{Y}$, are given by
\begin{equation}
\sigma_{XY}=\sqrt{\langle\hat{X}\hat{Y}\rangle - \langle\hat{X}\rangle\langle\hat{Y}\rangle}.
\end{equation}
To get the fluctuations of $\hat{X}$,  just use the above formula with  $\hat{X}=\hat{Y}$.
It can be shown that, in TDHF, the correlations at $t_1$ are obtained from
\begin{equation}
{\sigma_{XY}^2}^{(TDHF)}(t_1)=\text{Tr}\left\lbrace Y\rho(t_1) X[I-\rho(t_1)]\right\rbrace,\label{eq:sigmaTDHF}
\end{equation}
where $I$ is the identity, and $X$ and $Y$ are the matrices representing  the operators 
$\hat{X}$ and $\hat{Y}$.

One can also show that, in the TDRPA, this expression becomes
\begin{equation}
\label{eq:corr}
{\sigma_{XY}^2}^{(TDRPA)}(t_1)=\lim\limits_{\epsilon\rightarrow 0}\frac{\text{Tr}\left\lbrace\left[\rho(t_0)-\rho_X(t_0,\epsilon)\right]\left[\rho(t_0)-\rho_Y(t_0,\epsilon)\right]\right\rbrace}{2\epsilon^2},
\end{equation}
where the one-body density matrices $\rho_{X}(t,\epsilon)$ are solutions to the TDHF equation with the boundary condition
\begin{equation}
\label{eq:transform}
\rho_{X}(t_1,\epsilon)=e^{i\epsilon X} \rho(t_1) e^{-i\epsilon X}.
\end{equation}
To calculate the correlations, the state at $t_1$ is propagated backwards in time to the initial time $t_0$, explaining why the correlations at the time of interest, $t_1$, depends on the density matrices at the initial time, $t_0$.
Equations~(\ref{eq:sigmaTDHF}) and~(\ref{eq:corr}) are not equivalent. 
In fact, fluctuations computed from TDRPA are larger (in particular in DIC) than with TDHF, in better agreement with experiment. 

\begin{figure}[htbp] 
   \centering
   \includegraphics[width=4.7in]{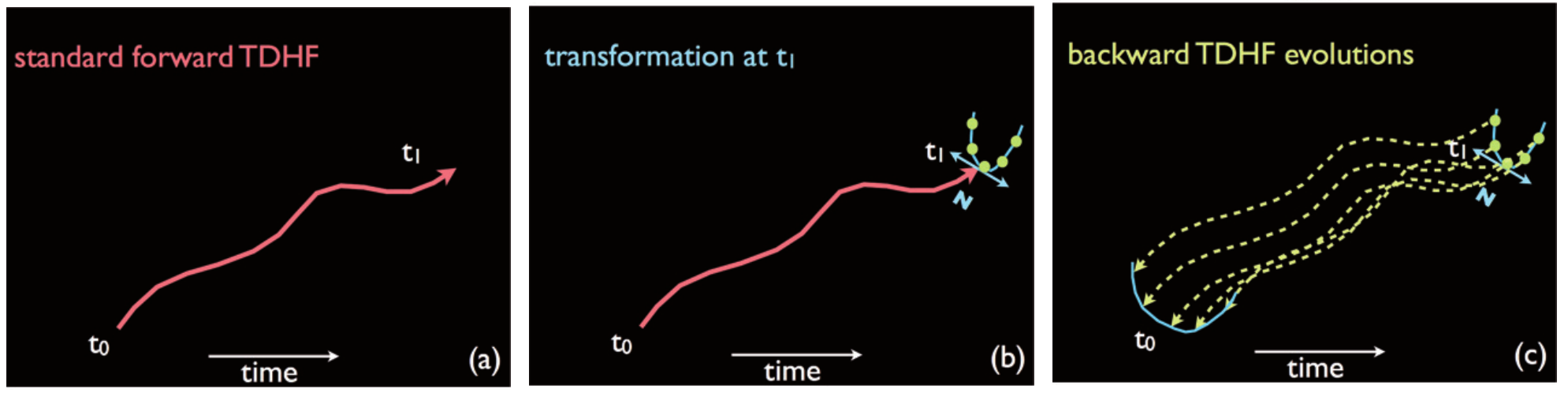} 
   \caption{Schematic representation of the TDRPA computational method to determine the fluctuations of $N$ at final time $t_1$. From (Simenel 2012).}
   \label{fig:TDRPA}
\end{figure}

To investigate fluctuations of particle number in outgoing fragments of deep-inelastic collisions, 
one can  use 
\begin{equation}
\hat{X}=\hat{N}_V = \sum_{sq}\int d\mathbf{r}~\hat{a}^\dagger(\mathbf{r}sq)\hat{a}(\mathbf{r}sq)~\Theta(\mathbf{r}),
\end{equation}
where $\Theta(\mathbf{r})=1$  in the volume $V$ containing the fragment, and $0$ elsewhere.
According to Eq.~(\ref{eq:transform}), the single-particle states (protons and/or neutrons) are transformed at $t_1$ as
\begin{equation}
\psi^X_i(\mathbf{r}s q, t_1;\epsilon) = e^{-i \epsilon \Theta(\mathbf{r})} \varphi_i(\mathbf{r}s q, t_1).
\end{equation}
These transformed states are then propagated backwards in time to the initial time $t_0$ for various (small) values of $\varepsilon$. 
The succession of forward and backward propagations is illustrated in Fig.~\ref{fig:TDRPA}.
To obtain the correlations (e.g., between proton and neutron numbers) and fluctuations, one then evaluates Eq.~(\ref{eq:corr}), which can be reduced to
\begin{equation}
\sigma_{XY}(t_0)=\sqrt{\frac{\eta_{00}(t_0)+\eta_{XY}(t_0)-\eta_{0X}(t_0)-\eta_{0Y}(t_0)}{2\epsilon^2}},
\end{equation}
where $\eta_{XX'}$ is
\begin{equation}
\eta_{XX'}=\sum_{i,j=1}^A\left|\left\langle\psi^X_i \big| \psi^{X'}_j \right\rangle\right|^2,
\end{equation}
with 
$X=0$ and $Y=0$ denoting the use of untransformed states, $\varphi_i$.

\subsubsection*{Acknowledgement}
This work has been supported by the Australian Research Council under Grant No. DP190100256.

\end{document}